# The MASSIVE survey – XIX. Molecular gas measurements of the supermassive black hole masses in the elliptical galaxies NGC 1684 and NGC 0997

Pandora Dominiak [1]★, Martin Bureau,[1]★ Timothy A. Davis,[2] Chung-Pei Ma,[3,4] Jenny E. Greene[5] and Meng Gu[6]

[1]*Sub-department of Astrophysics, Department of Physics, University of Oxford, Denys Wilkinson Building, Keble Road, Oxford OX1 3RH, UK*
[2]*Cardiff Hub for Astrophysics Research and Technology, School of Physics and Astronomy, Cardiff University, Queens Buildings, Cardiff CF24 3AA, UK*
[3]*Department of Physics, University of California, Berkeley, CA 94720, USA*
[4]*Department of Astronomy, University of California, Berkeley, CA 94720, USA*
[5]*Department of Astrophysical Sciences, Princeton University, Princeton, NJ 08544, USA*
[6]*The Observatories of the Carnegie Institution for Science, Pasadena, CA 91101, USA*

## ABSTRACT

Supermassive black hole (SMBH) masses can be measured by observing their dynamical effects on tracers, such as molecular gas. We present high angular resolution Atacama Large Millimeter/submillimeter Array observations of the $^{12}$CO(2–1) line emission of the early-type galaxies (ETGs) NGC 1684 and NGC 0997, obtained as part of the MASSIVE survey, a volume-limited integral-field spectroscopic study of the most massive local ETGs. NGC 1684 has a regularly rotating central molecular gas disc, with a spatial extent of $\approx$6 arcsec ($\approx$1.8 kpc) in radius and a central hole slightly larger than the expected SMBH sphere of influence. We forward model the data cube in a Bayesian framework with the KINEMATIC MOLECULAR SIMULATION (KINMS) code and infer a SMBH mass of $1.40^{+0.44}_{-0.39} \times 10^9$ M$_\odot$ ($3\sigma$ confidence interval) and an $F110W$-filter stellar mass-to-light ratio of $(2.50 \pm 0.05)$ M$_\odot$/L$_{\odot, F110W}$. NGC 0997 has a regularly rotating central molecular gas disc, with a spatial extent of $\approx$5 arcsec ($\approx$2.2 kpc) in radius and a partially filled central hole much larger than the expected SMBH sphere of influence, thus preventing a robust SMBH mass determination. With the same modelling method, we nevertheless constrain the SMBH mass to be in the range $4.0 \times 10^7$–$1.8 \times 10^9$ M$_\odot$ and the $F160W$-filter stellar mass-to-light ratio to be $(1.52 \pm 0.11)$ M$_\odot$/L$_{\odot, F160W}$. Both SMBH masses are consistent with the SMBH mass–stellar velocity dispersion ($M_{BH}$–$\sigma_e$) relation, suggesting that the overmassive SMBHs present in other very massive ETGs are fairly uncommon.

**Key words:** galaxies: elliptical and lenticular, cD – galaxies: individual: NGC 1684, NGC 0997 – galaxies: ISM – galaxies: kinematics and dynamics – galaxies: nuclei.

## 1 INTRODUCTION

Evidence has mounted over many years that each massive galaxy has a supermassive black hole (SMBH) at its centre. Although massive, the region an SMBH influences dynamically is very small compared to its host galaxy. Despite this, observations have also revealed that SMBH mass is tightly correlated with a number of host galaxy properties, such as bulge (and galaxy) mass and luminosity (e.g. Magorrian et al. 1998). The tightest correlation is that between SMBH mass ($M_{BH}$) and stellar velocity dispersion measured within the half-light (effective) radius ($\sigma_e$), commonly referred to as the $M_{BH}$–$\sigma_e$ relation (e.g. Kormendy & Ho 2013). These correlations imply that the SMBH somehow co-evolves with its host galaxy in a co-regulating manner, a finding that is vital to our current paradigm of galaxy evolution.

Studying these correlations requires reliable and accurate methods to measure SMBH masses. These include modelling the dynamics of integrated stellar populations, ionized gas, and megamasers (see compilations in, e.g. Kormendy & Ho 2013; McConnell & Ma 2013; van den Bosch et al. 2016). However, many of these methods have systematic weaknesses and/or a limited range of potential targets. Models of stellar kinematics rely on the absorption lines of the integrated stellar population spectra and are thus strongly affected by dust extinction. Stellar kinematics is thus primarily used in early-type galaxies (ETGs) with little dust. Ionized gas is particularly susceptible to non-gravitational forces and thus non-circular motions, which offer support against the gravity of the SMBHs and lead to SMBH mass underestimates. Maser dynamics have yielded the tightest constraints on SMBH masses so far, but only $\approx$ 3 per cent of local Seyfert 2 and low-ionization nuclear emission region galaxies have masers that can be used to estimate SMBH masses (van den

★ E-mail: pandora.dominiak@physics.ox.ac.uk (PD); martin.bureau@physics.ox.ac.uk (MB)

**Table 1.** Basic properties of the target galaxies.

| Target (1) | RA (2) | Dec. (3) | Hubble Type (4) | $D$ (Mpc) (5) | $\log(M_{H_2})$ ($M_\odot$) (6) | SFR ($M_\odot$ yr$^{-1}$) (7) | $M_K$ (mag) (8) | $\log(M_\star)$ ($M_\odot$) (9) | $\sigma_e$ (km s$^{-1}$) (10) | $\log(M_{BH})$ ($M_\odot$) (11) | $R_{SoI}$ (arcsec, pc) (12) |
|---|---|---|---|---|---|---|---|---|---|---|---|
| NGC 1684 | 04$^h$52$^m$31$^s$15 | −03°06′21″8 | E3 | 62.8 | 9.20 | 0.34 | −25.70 | 11.77 | 262 | 9.00 | 0.21, 63 |
| NGC 0997 | 02$^h$37$^m$14$^s$50 | +07°18′20″5 | E0 | 90.4 | 9.26 | 0.47 | −25.68 | 11.76 | 215 | 8.49 | 0.06, 28 |

*Note.*: Columns: (1) Target name. (2) Right ascension (J2000.0). (3) Declination (J2000.0). (4) Morphological classification. (5) Distance. (6) Total molecular gas mass. (7) Star formation rate. (8) Extinction-corrected total absolute *K*-band magnitude. (9) Total stellar mass estimated using $M_K$. (10) Average luminosity-weighted stellar velocity dispersion within one effective radius. (11) SMBH mass derived from the $M_{BH}$–$\sigma_e$ relation of van den Bosch (2016) using $\sigma_e$. (12) SMBH sphere of influence radius calculated using $\sigma_e$ and $M_{BH}$.

Bosch et al. 2016), all of which necessarily are relatively small active galaxies with correspondingly light SMBHs. There is thus a need for a unique method to measure SMBH masses, which can be used across all types of galaxies.

The general principle behind the aforementioned methods is to use a dynamical tracer to resolve the spatial scales on which a SMBH dominates the potential. In recent years a new method of SMBH mass measurement has been proposed and successfully developed – one which uses molecular gas (generally CO) as a dynamical tracer (e.g. Davis et al. 2013a). CO is an excellent dynamical tracer as it is unaffected by dust extinction and can be detected in all spiral galaxies and many ETGs, even those that are no longer forming stars. Consequently, as long as a suitable molecular gas reservoir exists within the gravitational sphere of influence of a SMBH and is kinematically undisturbed, CO dynamical modelling can be used in galaxies spanning a wide range of the $M_{BH}$–$\sigma_e$ relation.

In this work, we apply the CO dynamical modelling method to estimate the masses of the SMBHs in two galaxies from the MASSIVE survey, a volume-limited, multiwavelength, photometric, and spectroscopic survey of the most massive galaxies in the local Universe (Ma et al. 2014). Massive ETGs are the modern-day remnants of the earliest major star formation episodes in the history of our Universe, and are essential to understand cosmic evolution, galaxy formation, stellar populations, and SMBH formation. The MASSIVE sample includes 116 candidate galaxies in the northern sky with distances $D < 108$ Mpc and absolute integrated *K*-band magnitudes $M_K < -25.3$, corresponding to stellar masses $M_\star \gtrsim 10^{11.5}$ M$_\odot$. MASSIVE is designed to address a wide range of issues in elliptical galaxy formation and to date has investigated the photometric (Goullaud et al. 2018) and stellar kinematic (Veale et al. 2017a, b, 2018; Ene et al. 2018; Ene et al. 2019, 2020) properties of its sample galaxies, the break in the high-mass end of the Tully–Fisher relation (Davis et al. 2015), the properties of hot and warm ionized gas (Goulding et al. 2016; Pandya et al. 2017), and the primary drivers of the variations of the global stellar initial mass function (IMF; Gu et al. 2022), all in the most massive ETGs. A systematic study of the molecular gas content of 67 representative MASSIVE galaxies reports a 25 per cent detection rate of CO line emission to a sensitivity of a molecular-to-stellar mass fraction of $\approx$0.1 per cent (Davis et al. 2019).

Another primary goal of the MASSIVE survey is to detect and measure the dynamical masses of central SMBHs within this homogeneously selected galaxy sample. Thus far only 12 of the MASSIVE galaxies have had their SMBH mass estimates from direct dynamical modelling published. Most of these measurements are obtained from spatially resolved stellar kinematics and axisymmetric Schwarzschild modelling of stellar orbits: in two brightest cluster galaxies in rich clusters, NGC 4889 and NGC 3842 (McConnell et al. 2011, 2012); two brightest galaxies in galaxy groups, NGC 1600 (Thomas et al. 2016) and NGC 7619 (Rusli et al. 2013); and three Virgo cluster galaxies, M87 (Gebhardt et al. 2011), NGC 4649 (Shen & Gebhardt 2010), and NGC 4472 (Rusli et al. 2013). More recently, the MASSIVE team has relaxed the axisymmetric assumption and performed triaxial orbit modelling to obtain new SMBH masses for two fast-rotating MASSIVE ETGs, NGC 1453 and NGC 2693 (Liepold et al. 2020; Pilawa et al. 2022; Quenneville, Liepold & Ma 2022). A new determination of the M87 SMBH mass based on the same triaxial orbit code, TRIOS, is reported in Liepold, Ma & Walsh (2023). Additionally, three SMBH masses in MASSIVE galaxies have been determined from dynamical modelling of CO kinematics: NGC 315 (Boizelle et al. 2021), NGC 383 (North et al. 2019), and NGC 7052 (Smith et al. 2021).

This paper presents observations and kinematic modelling of the $^{12}$CO(2–1) line emission of two additional MASSIVE ETGs observed at high angular resolution with Atacama Large Millimeter/submillimeter Array (ALMA): NGC 1684 and NGC 0997. SMBH mass measurements using the same technique have been also presented for a number of other nearby galaxies (Davis et al. 2013b, 2017a, b, 2020; Barth et al. 2016; Onishi et al. 2017; Boizelle et al. 2019; Ruffa et al. 2019, 2023; Smith et al. 2019; Kabasares et al. 2022).

The paper is structured as follows. In Section 2 we summarize the main properties of the two targets. The ALMA data used in this work, their reduction, and the properties of the CO cube and the continuum emission are described in Section 3. The molecular gas dynamical modelling is described in Section 4. We present the results in Section 5 and discuss them in Section 6 before summarizing and concluding in Section 7. In Appendix A we provide a comparison between the data and best-fitting model CO moment maps and in Appendix B we provide first-moment residuals (model-data).

## 2 TARGETS

The basic properties of our targets are summarized in Table 1.

### 2.1 NGC 1684

NGC 1684 is an E3 elliptical galaxy located at 04$^h$52$^m$31$^s$15, −03°06′21″8 (J2000.0). Throughout this paper we adopt the distance used in the MASSIVE survey, $D = 62.8 \pm 2.3$ Mpc, obtained through surface brightness fluctuation (Jensen et al. 2021). At this distance, 1 arcsec corresponds to ≈304 pc. NGC 1684 is in a group of 11 galaxies according to the Two Micron All Sky Survey (Skrutskie et al. 2006) Redshift Survey (2MRS; Crook et al. 2007), of which it is the brightest. There is a much smaller elliptical galaxy-like object just ≈6 arcsec west of NGC 1684, most likely a low-mass galaxy in the process of being accreted (see Fig. 1).

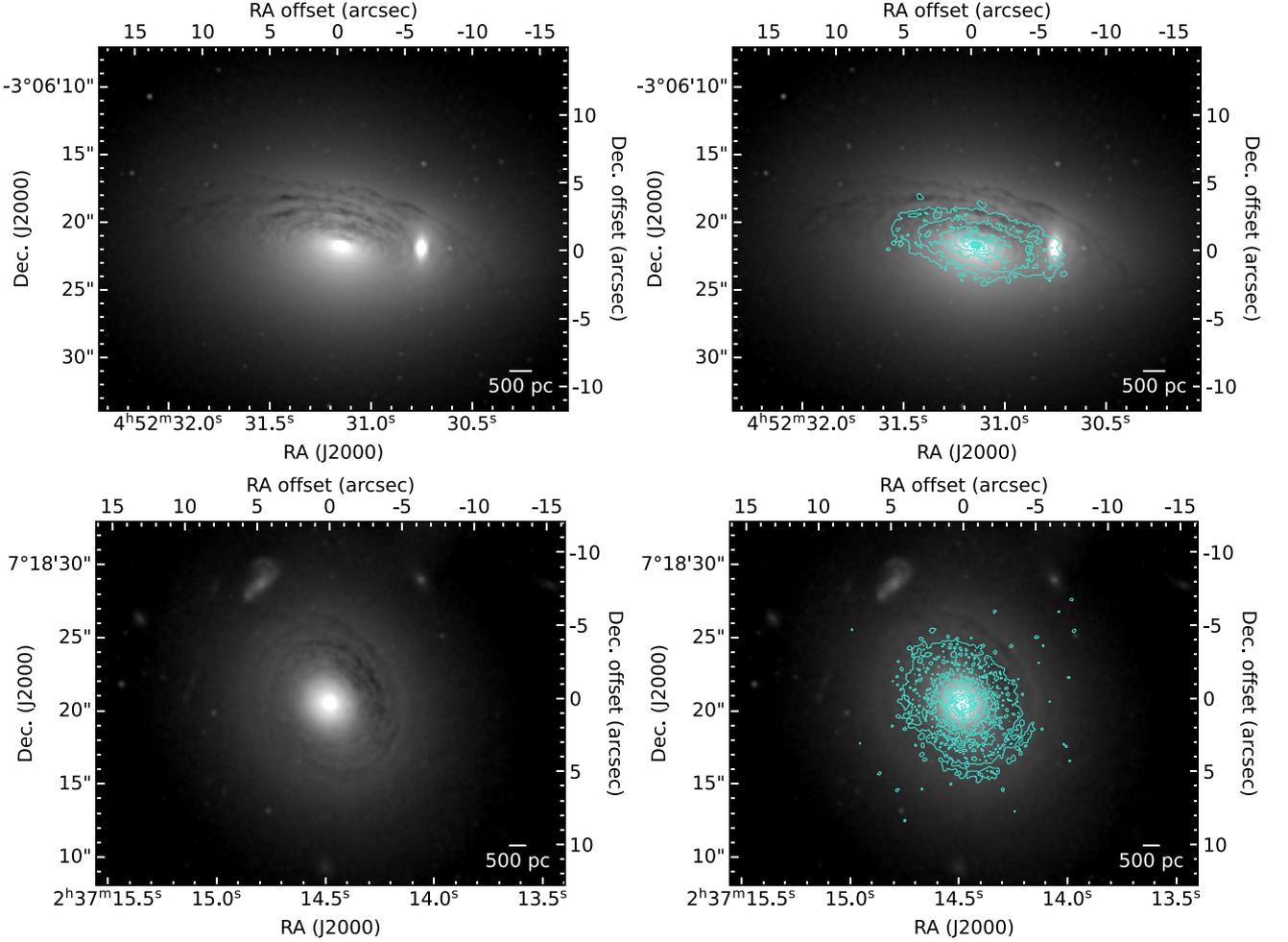

**Figure 1.** **Top:** Unsharp-masked *HST* Wide Field Camera 3 (WFC3) *F*110*W* filter (*J* band) image of NGC 1684 (left), overlaid with the $^{12}$CO(2–1) integrated intensity contours (cyan) from our ALMA observations (right). **Bottom:** Unsharp-masked *HST* WFC3 *F*160*W* filter (*H* band) image of NGC 0997 (left), overlaid with the $^{12}$CO(2–1) integrated intensity contours (cyan) from our ALMA observations (right).

Davis et al. (2015) estimated the inclination of the galaxy to be $42 \pm 5°$, but as neither dust lane nor interferometric observations of the source existed at the time, this was estimated using the axial ratio from the NASA/IPAC Extragalactic Database (NED). *Hubble Space Telescope* (*HST*) optical images reveal a prominent dust disc, with dust lanes to the north-west of the galaxy centre and an unobscured nucleus (Fig. 1). The dust disc has a semimajor axis of 24.5 arcsec and a semiminor axis of 9.6 arcsec, yielding an inclination of $\approx 67°$, clearly suggesting that the previous estimate was too low.

NGC 1684 hosts a significant molecular gas reservoir, with a total H$_2$ mass of $1.58^{+0.08}_{-0.07} \times 10^9$ M$_\odot$, and has a total star formation rate (SFR) of $0.34 \pm 0.03$ M$_\odot$ yr$^{-1}$ derived from corrected 22 μm luminosities (Davis et al. 2015). It has an integrated absolute *K*-band magnitude of $-25.70$ (Quenneville et al. 2023), which results in a total stellar mass of $5.86 \times 10^{11}$ M$_\odot$ (calculated using equation 2 of Cappellari 2013).

NGC 1684 does not have a direct SMBH mass measurement, but an estimate can be obtained using the $M_{BH}$–$\sigma_e$ relation of van den Bosch (2016). Assuming $\sigma_e = 262$ km s$^{-1}$ (Veale et al. 2018), a SMBH mass of $\approx 1.0 \times 10^9$ M$_\odot$ is expected. The radius of the sphere of influence ($R_{SoI}$) of a SMBH quantifies the spatial scale over which the gravitational potential of a SMBH is dominant. One estimate for the sphere of influence is $R_{SoI} \equiv G M_{BH}/\sigma_e^2$, where $G$ is the gravitational constant, thus yielding $R_{SoI} \approx 63$ pc (or $\approx 0.21$ arcsec) for NGC 1684.

### 2.2 NGC 0997

NGC 0997 is an E0 elliptical located at 02$^h$37$^m$14$^s$.50, +07°18′20″.5 (J2000.0). Throughout this paper we adopt the distance used in the MASSIVE survey, $D = 90.4$ Mpc (Ma et al. 2014). At this distance 1 arcsec corresponds to $\approx 438$ pc. NGC 0997 is located in a group of three galaxies (according to 2MRS; Crook et al. 2007), of which it is the brightest, and the galaxy distance was obtained by correcting for local peculiar velocities using group-corrected redshift distances (Ma et al. 2014).

*HST* optical images reveal a prominent dust disc, with dust lanes on all sides of the galactic centre and an unobscured nucleus (Fig. 1). There is a small and very faint galaxy-like object $\approx 9$ arcsec north-north-east of NGC 0997. Slightly further out, $\approx 16$ arcsec north of NGC 0997 (not visible in Fig. 1), there is another small object (PGC 200205) about which not much is known, but due to its apparent size it is most likely also a background galaxy.

Davis et al. (2015) estimated the inclination of the galaxy to be $34 \pm 5°$, consistent with that derived in this paper (Section 5). NGC 0997 hosts a molecular gas reservoir, with a total H$_2$ mass

**Table 2.** ALMA observations properties.

| Target | Track | Date | Array | Baseline range | On-source time (s) | MRS (arcsec, kpc) | FoV (arcsec, kpc) | Calibration |
|---|---|---|---|---|---|---|---|---|
| (1) | (2) | (3) | (4) | (5) | (6) | (7) | (8) | (9) |
| NGC 1684 | uid_A002_Xb12f3b_X825a | 2016-03-31 | 12-m | 15–453 m | 484 | 6.8, 2.1 | 24.8, 7.5 | Pipeline |
|  | uid_A002_Xbb30f0_X1ae5 | 2016-11-29 | 12-m | 15–704 m | 302 | 4.3, 1.3 | 24.8, 7.5 | Pipeline |
|  | uid_A002_Xc2f3fc_X248e | 2017-08-02 | 12-m | 54 m–3.3 km | 998 | 1.3, 0.4 | 24.8, 7.5 | Pipeline |
| NGC 0997 | uid_A002_Xb6e98e_X5d28 | 2016-08-19 | 12-m | 15 m–1.5 km | 759 | 3.9, 1.7 | 24.9, 10.9 | Pipeline |
|  | uid_A002_Xbb154b_X1dbd | 2016-11-26 | 12-m | 15–704 m | 423 | 4.3, 1.9 | 24.9, 10.9 | Pipeline |
|  | uid_A002_Xbb154b_X1dbd | 2017-08-22 | 12-m | 21 m–3.7 km | 1361 | 1.5, 0.6 | 24.9, 10.9 | Pipeline |

*Note.* Columns: (1) Target name. (2) Track ID. (3) Observations dates. (4) Type of ALMA antennae (ALMA array). (5) Minimum and maximum baseline lengths. (6) Total integration time on-source. (7) Maximum recoverable scale (MRS), i.e. largest angular scale structure that can be recovered with the given array configuration. (8) Field of view (FoV), i.e. primary beam full width at half-maximum. (9) Calibration method.

of $1.82^{+0.13}_{-0.12} \times 10^9$ M$_\odot$, and has a total SFR of $0.47 \pm 0.04$ M$_\odot$ yr$^{-1}$, determined as above. It has an integrated absolute *K*-band magnitude of $-25.68$ (Quenneville, Liepold & Ma 2022), which results in a total stellar mass of $5.74 \times 10^{11}$ M$_\odot$, calculated as above.

NGC 0997 does not have a direct SMBH mass measurement. Assuming $\sigma_e = 215$ km s$^{-1}$ (Davis et al. 2019) and the $M_{\rm BH}$–$\sigma_e$ relation of van den Bosch (2016), a SMBH mass of $\approx 3.0 \times 10^8$ M$_\odot$ is expected, yielding $R_{\rm SoI} \approx 28$ pc ($\approx 0.06$ arcsec).

## 3 ALMA OBSERVATIONS

Observations of the $^{12}$CO(2–1) emission line using the 12-m ALMA array were obtained as part of projects 2015.1.00187 (PI: Davis) and 2016.1.00683 (PI: Ma). The NGC 1684 data are comprised of three tracks taken on 2016 March 31, 2016 November 29, and 2017 August 2, for a total of 30 min on source. Baselines span 15.1 m to 3.3 km, yielding sensitivity to angular scales from 0.1 to 22 arcsec and a maximum recoverable scale of 6.8 arcsec. The NGC 0997 data are also comprised of three tracks taken on 2016 August 19, 2016 November 26, and 2017 August 22, for a total of 42 min on source. Baselines span 15.1 m to 3.7 km, yielding sensitivity to angular scales of 0.09 to 22 arcsec and a maximum recoverable scale of 4.3 arcsec. The properties of the observations are listed in Table 2.

ALMA Band 6 was used for all observations. All of the six tracks had a total of four spectral windows. For all tracks, one spectral window was used to map the molecular gas and was centred on the redshifted frequency of the $^{12}$CO(2–1) line (rest frequency $\nu_{\rm rest} = 230.5380$ GHz), with a bandwidth of 1.875 GHz ($\approx 2440$ km s$^{-1}$). For one track for each galaxy this spectral window was subdivided into 3840 channels of width $\approx 488$ kHz ($\approx 0.64$ km s$^{-1}$), while for the other two tracks for each galaxy this spectral window was subdivided into 1920 channels of width $\approx 977$ kHz ($\approx 1.27$ km s$^{-1}$). For all tracks, the remaining three spectral windows were used to map the continuum, each with a bandwidth of 2 GHz ($\approx 2600$ km s$^{-1}$). For one track for each galaxy these spectral windows were subdivided into 128 channels of width $\approx 16$ MHz ($\approx 20$ km s$^{-1}$), while for the other two tracks for each galaxy these spectral windows were subdivided into 240 channels of width $\approx 8$ MHz ($\approx 10$ km s$^{-1}$). For each track, bright quasars were used as bandpass, flux, phase, and water vapour calibrators.

The data for both galaxies were calibrated using the standard ALMA pipeline and were concatenated using the default weighting, using the COMMON ASTRONOMY SOFTWARE APPLICATIONS (CASA) package (McMullin et al. 2007).

**Table 3.** CO data cube properties.

| Target | Image property | Value |
|---|---|---|
| NGC 1684 | Spatial extent (pix) | 1024 × 1024 |
|  | Spatial extent (arcsec) | 51.2 × 51.2 |
|  | Spatial extent (kpc) | 15.6 × 15.6 |
|  | Pixel scale (arcsec pix$^{-1}$) | 0.05 |
|  | Pixel scale (pc pix$^{-1}$) | 15.2 |
|  | Velocity range (km s$^{-1}$) | 3950 – 4850 |
|  | Channel width (km s$^{-1}$) | 10 |
|  | RMS noise (mJy beam$^{-1}$ channel$^{-1}$) | 0.53 |
|  | Number of constraints | 194,887 |
|  | Synthesized beam (arcsec) | 0.21 × 0.14 |
|  | Synthesized beam (pc) | 64 × 43 |
| NGC 0997 | Spatial extent (pix) | 1024 × 1024 |
|  | Spatial extent (arcsec) | 51.2 × 51.2 |
|  | Spatial extent (kpc) | 22.4 × 22.4 |
|  | Pixel scale (arcsec pix$^{-1}$) | 0.05 |
|  | Pixel scale (pc pix$^{-1}$) | 21.9 |
|  | Velocity range (km s$^{-1}$) | 6000 – 6700 |
|  | Channel width (km s$^{-1}$) | 5 |
|  | RMS noise (mJy beam$^{-1}$ channel$^{-1}$) | 0.51 |
|  | Number of constraints | 145,141 |
|  | Synthesized beam (arcsec) | 0.17 × 0.15 |
|  | Synthesized beam (pc) | 77 × 64 |

### 3.1 Line emission

For each galaxy, the continuum spectral windows and line-free channels of the line spectral window were linearly fit and the fit subtracted from the *uv*-plane data using the CASA uvcontsub task. The continuum-subtracted *uv* data were then imaged and cleaned in regions of source emission to a threshold equal to the root-mean-square (RMS) noise of the dirty channels (in regions free of line emission), using the cube mode of the CASA task tclean and Briggs weighting with a robust parameter of 0.5. We adopt a channel width of 10 km s$^{-1}$ for NGC 1684, the typical velocity resolution used in molecular gas determinations of SMBH masses. We adopt a channel width of 5 km s$^{-1}$ for NGC 0997, smaller than typical, to aid SMBH mass detection (Davis 2014), but still much larger than the raw channel width of the observations. The resulting data cubes have a synthesized beam of 0.21 arcsec × 0.14 arcsec (64 × 43 pc$^2$; sampled with $\approx$3–5 spaxels linearly) with an RMS noise of 0.53 mJy beam$^{-1}$ channel$^{-1}$ for NGC 1684 and 0.17 arcsec × 0.15 arcsec (77 × 64 pc$^2$) with an RMS noise of 0.51 mJy beam$^{-1}$ channel$^{-1}$ for NGC 0997. All the properties of the resulting data cubes are summarized in Table 3.

The zeroth (integrated-intensity), first (intensity-weighted mean line-of-sight velocity), and second (intensity-weighted line-of-sight velocity dispersion) moment maps of NGC 1684 and NGC 0997 are shown in Figs 2 and 3, respectively, along with kinematic major-axis position–velocity diagrams (PVDs). The maps were created using the masked-moment technique (Dame 2011) implemented in the PYMAKEPLOTS package.[1] First, a copy of the CO data cube without primary-beam correction is boxcar-smoothed in the spatial and spectral dimensions, with a spatial kernel of 1.5 times the width of the synthesized beam and a spectral kernel of four times the channel width. A binary mask is then created by clipping the smoothed data cube at a threshold of five times the RMS noise of the line-free channels of the smoothed data cube. The moment maps and PVDs are then created by applying this mask to the unsmoothed primary beam-corrected (i.e. the original) data cube. The PVDs were constructed by summing pixels in each masked cube within a pseudo-slit 5 pixels wide oriented along the kinematic position angle of 258.1° for NGC 1684 and 215.2° for NGC 0997, determined by our kinematic modelling results in Section 5.

In NGC 1684, we clearly detect a disc of molecular gas in regular rotation and coincident with (but smaller than) the dust disc (see Fig. 1), with a central hole of $\approx 0.35$ arcsec in radius. We also detect a disc of molecular gas in regular rotation in NGC 0997, again coincident with the dust disc and with a semi-annular central hole which extends to a maximal radius of $\approx 0.5$ arcsec. Despite this, some molecular gas extends past the hole towards the centre of the galaxy. Such holes are frequently detected in galaxies where the molecular gas has been sufficiently well spatially resolved, and may be due to a number of processes [e.g. shear and/or tidal disruption of gas clouds (Liu et al. 2021; Smith et al. 2021); photodissociation by a hard radiation field due to an active galactic nucleus (AGN; Visser, van Dishoeck & Black 2009); thermal effects from an accretion disc reducing the fraction of gas emitting in the low-$J$ transitions (Wada et al. 2018)]. Likewise, the asymmetric hole in NGC 0997 can be due to a number of reasons, e.g. photodissociation due to an AGN with asymmetric jets, asymmetric dust shielding of thermal effects, and/or molecular gas not yet being fully settled.

For NGC 1684, the hole is slightly larger than the predicted radius of the SMBH sphere of influence, and thus does not allow for the innermost parts of the potential to be probed. Consequently, no Keplerian rise is present within $R_{\rm SoI}$. SMBH mass measurements are still possible in such instances, as the presence of the SMBH will also enhance the gas rotation velocities at radii beyond the formal SMBH sphere-of-influence. However, the effect of the SMBH then becomes increasingly degenerate with that of the stars. For NGC 0997, the semi-annular hole poses a unique case. As the molecular gas on one side of the galaxy extends to its centre, a small enhancement of velocities can be seen on one side of the PVD at $\approx 0.2$ arcsec, which is partly within $R_{\rm SoI}$. However, the hole prevents this enhancement from being detected on the other side. In cases where there is no clear Keplerian rise within $R_{\rm SoI}$, it may not be possible to obtain a robust SMBH mass, but a SMBH mass upper limit can nevertheless usually be estimated.

Fig. 4 shows the integrated $^{12}$CO(2–1) spectra of NGC 1684 and NGC 0997, both with the classic double-horn shape of a rotating disc. These spectra are extracted from a 16 arcsec × 8 arcsec ($\approx 4860 \times 2430$ pc$^2$) region around the centre of NGC 1684 (that includes all detected emission), with an integrated line flux (measured within the mask defined in Section 3.1) of $99.9 \pm 3.9$ Jy km s$^{-1}$, and a 16 arcsec × 16 arcsec ($\approx 7000 \times 7000$ pc$^2$) region around the centre of NGC 0997, with an integrated line flux of $52.8 \pm 1.5$ Jy km s$^{-1}$. Fig. 4 also shows the integrated spectra of the $^{12}$CO(2–1) (grey shading) and $^{12}$CO(1–0) (black dashed line) lines obtained from the Institut de Radio Astronomie Millimétrique (IRAM) single-dish 30-m telescope observations reported by Davis et al. (2015). The 30-m telescope $^{12}$CO(2–1) integrated intensity is smaller than that measured by ALMA for both galaxies, likely a consequence of flux omission due to the smaller primary beam ($\approx 11$ arcsec) of the 30-m telescope, as well as potential pointing errors. The lopsided spectrum of NGC 1684 is a classic sign of the latter, suggesting the 30-m telescope was pointing slightly to the north-east of the galaxy centre.

As we do not have $^{12}$CO(1–0) observations ourselves, we adopt the total molecular gas masses (including contributions from heavy elements) estimated from the above $^{12}$CO(1–0) 30-m telescope observations of Davis et al. (2015) in the standard manner [i.e. using equation 3 of Bolatto, Wolfire & Leroy 2013 and a standard CO-to-molecule conversion factor $\alpha_{\rm CO(1-0)} = 4.3$ M$_\odot$ pc$^{-2}$ (K km s$^{-1}$)$^{-1}$], yielding $(1.58 \pm 0.07) \times 10^9$ M$_\odot$ for NGC 1684 and $(1.82 \pm 0.13) \times 10^9$ M$_\odot$ for NGC 0997. From our ALMA $^{12}$CO(2–1) and the 30-m telescope $^{12}$CO(1–0) integrated line intensities, we derive $^{12}$CO(2–1)-to-$^{12}$CO(1–0) flux ratios (expressed in Jy km s$^{-1}$) $R_{21} = 2.97$ for NGC 1684 and $R_{21} = 2.82$ for NGC 0997.

### 3.2 Continuum emission

An image of the 236.1 GHz (1.27 mm) continuum emission of NGC 1684 was created using the CASA task tclean in multifrequency synthesis mode and Briggs weighting with a robust parameter of 0.5. The line-free channels of the line spectral window were used as well as the continuum spectral windows. A small central source was detected and fitted with a Gaussian function using the CASA task imfit, revealing a spatially resolved source with an integrated flux density of $22.9 \pm 0.5$ mJy. All the properties of the NGC 1684 continuum image and the detected source are listed in Table 4. An image of the 234.6 GHz (1.28 mm) continuum emission of NGC 0997 was created in an analogous manner. A small spatially resolved continuum source was again detected, with an integrated flux density of $6.58 \pm 0.08$ mJy. All the properties of the NGC 0997 continuum image and the detected source are listed in Table 5.

## 4 DYNAMICAL MODELLING

### 4.1 Model overview

Dynamical modelling of each galaxy was carried out by fitting the observed ALMA data cube discussed in the previous section (both the molecular gas distribution and its kinematics) with a model. Each model data cube is created using the Python version of KINEMATIC MOLECULAR SIMULATION[2] (KINMS; Davis et al. 2013a). As described in Section 4.2, this requires a parametric description (with a number of free parameters) of the surface density profile of the tracer (here $^{12}$CO(2–1)), to create a set of particles that are then appropriately projected (i.e. positioned) within the model cube.

Prior to constraining the SMBH (whose mass is another parameter of the model), we must calculate the stellar contribution to the total gravitational potential. As described in Section 4.3, this is achieved by first parametrizing the stellar light distribution using a multi-Gaussian expansion[3] (MGE; Emsellem, Monnet & Bacon 1994; Cappellari 2002) of an optical image, and then converting this to a stellar mass distribution using a free radially constant stellar mass-to-light ratio $M/L$, simultaneously yielding the circular velocity curve of the model.

---

[1] https://github.com/TimothyADavis/pymakeplots

[2] https://github.com/TimothyADavis/KinMS_fitter

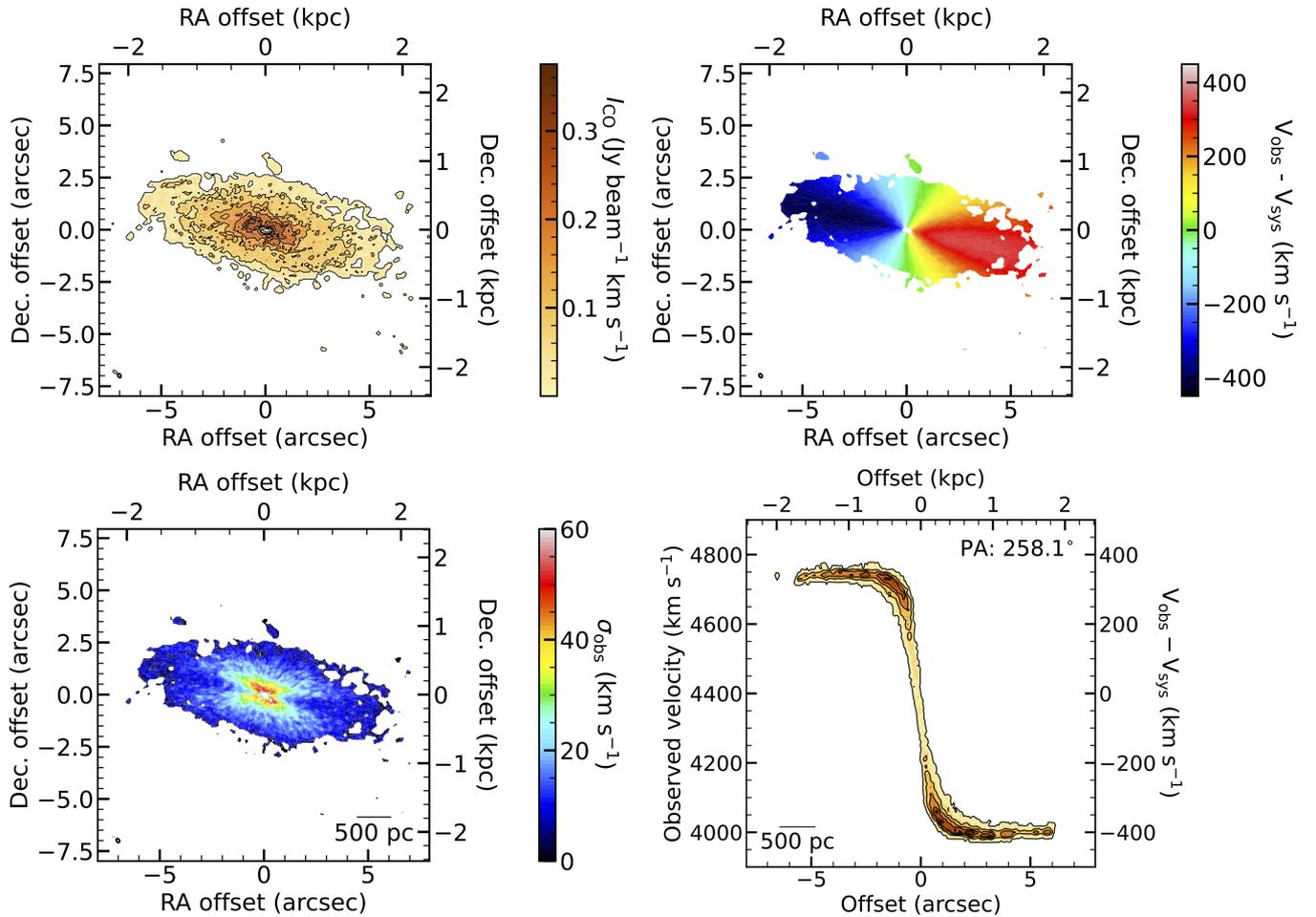

**Figure 2.** $^{12}$CO(2–1) emission properties of NGC 1684 derived from our ALMA data. **Top-left:** zeroth-moment (integrated-intensity) map. **Top-right:** first-moment (intensity-weighted mean line-of-sight velocity) map. **Bottom-left:** second-moment (intensity-weighted line-of-sight velocity dispersion) map. **Bottom-right:** kinematic major-axis PVD. The synthesized beam is shown in the bottom-left corner of each map as an open ellipse. A scale bar is shown in the bottom-right corner of each map and the bottom-left corner of the PVD. We note the presence of a small central hole in the molecular gas distribution. Likely as a result, there is no evidence of a central (Keplerian) rise of the rotation velocities at small radii.

Finally, we must compare the ALMA data cube against model data cubes. As described in Section 4.4, we use a Markov chain Monte Carlo (MCMC) method that fully samples the parameter space to estimate the posterior distributions and hence the uncertainties on the best-fitting free parameters.

### 4.2 Gas distributions

As a part of our modelling approach, the molecular gas distribution of each galaxy must be quantified. A thin exponential disc is usually appropriate for ETGs (Davis et al. 2013a), but at the spatial resolutions achieved by ALMA central holes, depressions and/or complex features are often detected, as is the case with both NCG 1684 and NGC 0997.

In the case of NGC 1684, we therefore adopt a 2D axisymmetric exponential surface density profile with an inner truncation at a given radius:

$$I(R) \propto \begin{cases} 0 & R \leq R_{\text{hole}}, \\ e^{-R/R_0} & R > R_{\text{hole}}, \end{cases} \quad (1)$$

where $R$ is the galactocentric radius, $R_0$ the exponential scale length, and $R_{\text{hole}}$ is the truncation radius, the latter two being free parameters of our model. The entire model data cube is also scaled to match the integrated $^{12}$CO(2–1) flux, requiring another free parameter.

For completeness, we considered alternative surface density profiles. We tried an exponential profile at large radii, truncated at an inner radius but extending inwards as a half-Gaussian function. This model seemed more appropriate for NGC 1684 considering that it has some form of depression around the hole, but the fitting algorithm could not converge, most likely because the inner half-Gaussian parameters are not strongly constrained due to the limited spatial resolution of the observations. Other forms of extension inwards from the truncation fared no better.

In the case of NGC 0997, all fits using any combination of parametric functions failed to adequately reproduce the observed gas distribution. However, there is no intrinsic need to specify the distribution analytically, and in cases where the gas is particularly morphologically complex, as in NGC 0997, we can use the observed gas distribution instead. We implemented this using the SKYSAMPLER[4] tool described in Smith et al. (2019). First, we obtained the kinematics-independent distribution of the gas by integrating the point source model (i.e. the clean components) of the gas distribution along the velocity axis. We then sampled this with a large number of particles. Assuming a certain position angle and inclination for the galaxy, the particle positions can then be deprojected into their intrinsic positions in the galaxy plane.

[3] https://pypi.org/project/mgefit/

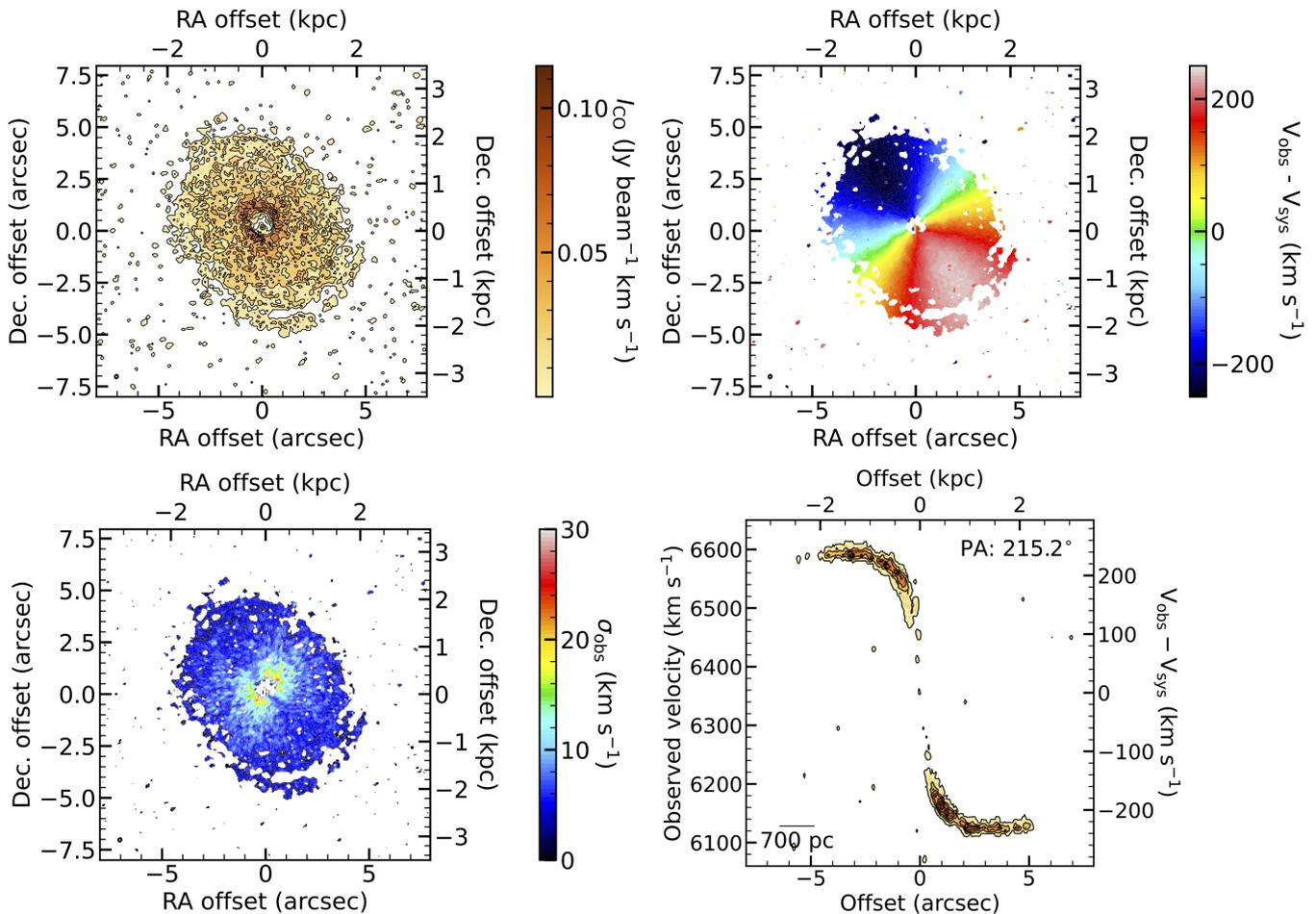

**Figure 3.** As Fig. 2 but for NGC 0997. We note the presence of a small, asymmetric central hole, as well as a faint secondary peak in the very centre of the molecular gas distribution. Likely as a result of the hole, there is no evidence of a central (Keplerian) rise of the rotation velocities at small radii. However, as a result of the secondary peak, we do see an asymmetric enhancement of velocities at the very centre.

Additional to the molecular gas distribution, our model also includes a gas velocity dispersion ($\sigma_{\rm gas}$), which is assumed to be spatially constant and small relative to the rotational velocities, which is typically the case in ETGs (often <10 km s$^{-1}$; e.g. Davis et al. 2017a, Smith et al. 2019).

### 4.3 Stellar distributions

To estimate the SMBH mass of each galaxy accurately, the stellar mass contribution must first be quantified and accounted for. The kinematics are in principle also affected by dark matter, but at the small radii probed here the dark matter contribution will be assumed to be negligible (it would in any case be largely degenerate with that of the stars).

We parametrize the stellar light distribution of NGC 1684 using an MGE model of a *HST* Wide Field Camera 3 (WFC3) image in the *F*110*W* filter (*J* band) (GO-14219, PI: Blakeslee) with a total exposure time of 2496 s, and similarly for NGC 0997, using a *HST* WFC3 image in the *F*160*W* filter (*H* band) (GO-15909, PI: Boizelle) with a total exposure time of 996 s. In each case this is the longest wavelength *HST* image available, to minimize dust extinction. We obtain the point spread function (PSF) of each filter using TINYTIM (Krist, Hook & Stoehr 2011) and parametrize each by fitting a circular MGE model, listed in Table 6.

Prior to comparison with the observed light distribution, the MGE surface brightness is convolved with the PSF calculated above in order to account for instrumental effects. The 2D projection of the stellar light distribution captured by each image is then parametrized by the MGE as a sum of Gaussians, each with a central surface brightness $I$, a width $\sigma$, and an axial ratio $q$. The MGE central surface brightnesses can be converted into luminosities, and in the AB magnitude system we adopt a zero-point of 26.82 mag and a Solar absolute magnitude of 4.52 mag for the *F*110*W* filter, and a zero-point of 25.94 mag and a Solar absolute magnitude of 4.60 mag for the *F*160*W* filter (see Sahu, Anderson & Baggett 2021 and Willmer 2018, respectively). We also adopt galactic extinctions of 0.051 mag

---
[4] https://github.com/Mark-D-Smith/KinMS-skySampler

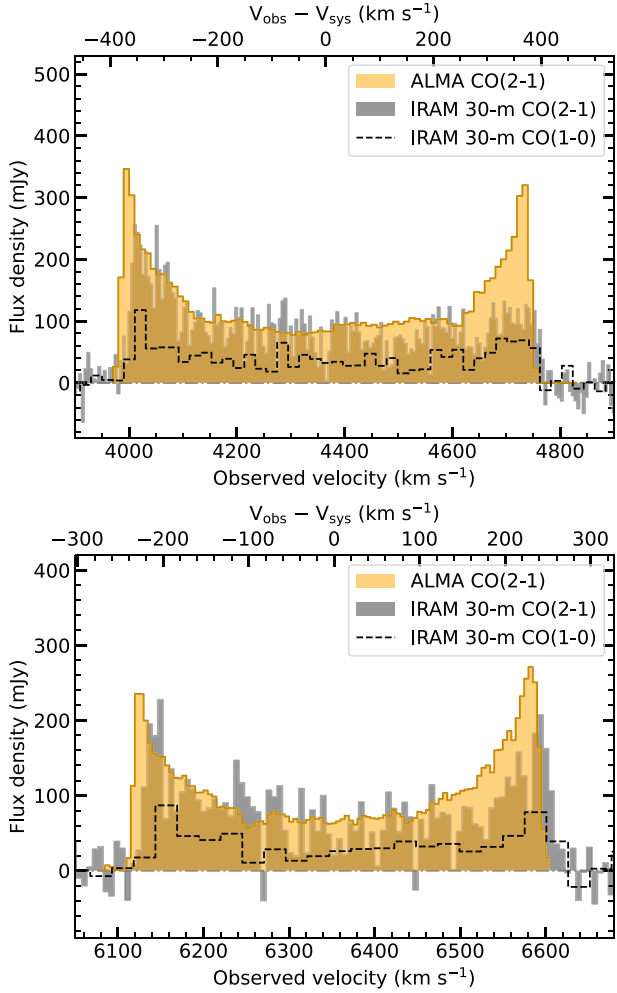

**Figure 4.** **Top:** $^{12}$CO(2–1) integrated spectrum of NGC 1684 derived from our ALMA data (orange shading), overlaid on the IRAM 30-m telescope $^{12}$CO(2–1) (grey shading) and $^{12}$CO(1–0) (black dashed line) observations of Davis et al. (2015). The ALMA aperture is a 16 arcsec × 8 arcsec ($\approx$4860 × 2430 pc$^2$) region centred on the galaxy centre, which includes all detected emission. **Bottom:** As the top panel, but for NGC 0997, with an ALMA aperture of 16 arcsec × 16 arcsec ($\approx$7000 × 7000 pc$^2$). The grey dash–dotted line in both figures indicates the zero flux level. Both ALMA spectra show the characteristic double-horn shape of a rotating disc.

**Table 4.** Parameters of the NGC 1684 continuum image and the detected continuum source.

| Image property | Value |
| --- | --- |
| Image (pix) | 512 × 512 |
| Image (arcsec) | 10.2 × 10.2 |
| Image size (kpc) | 3.10 × 3.10 |
| Pixel scale (arcsec pix$^{-1}$) | 0.02 |
| Pixel scale (pc pix$^{-1}$) | 6.08 |
| RMS noise (mJy beam$^{-1}$) | 0.12 |
| Synthesized beam (arcsec) | 0.18 × 0.11 |
| Synthesized beam (pc) | 55 × 33 |
| Source property | Value |
| Right ascension | 04$^h$52$^m$31$^s$.1410 ± 0$^s$.0002 |
| Declination | −03°06′21″.624 ± 0″.002 |
| Integrated flux (mJy) | 22.9 ± 0.5 |
| Deconvolved size (marcsec) | (267 ± 9) × (186 ± 8) |
| Deconvolved size (pc) | (81 ± 3) × (57 ± 2) |

**Table 5.** Parameters of the NGC 0997 continuum image and the detected continuum source.

| Image property | Value |
| --- | --- |
| Image size (pix) | 1024 × 1024 |
| Image size (arcsec) | 51.2 × 51.2 |
| Image size (kpc) | 22.44 × 22.44 |
| Pixel scale (arcsec pix$^{-1}$) | 0.05 |
| Pixel scale (pc pix$^{-1}$) | 21.9 |
| RMS noise (mJy beam$^{-1}$) | 0.04 |
| Synthesized beam (arcsec) | 0.15 × 0.12 |
| Synthesized beam (pc) | 66 × 53 |
| Source property | Value |
| Right ascension | 02$^h$37$^m$14$^s$.47984 ± 0$^s$.00007 |
| Declination | +07°18′20″.4706 ± 0″.0006 |
| Integrated flux (mJy) | 6.58 ± 0.08 |
| Deconvolved size (marcsec) | (195 ± 4) × (114 ± 3) |
| Deconvolved size (pc) | (85 ± 2) × (50 ± 1) |

**Table 6.** Parameters of the *HST* WFC3 PSFs.

| Filter | $G_k$ | $\sigma_k$ (arcsec) |
| --- | --- | --- |
| (1) | (2) | (3) |
| *F*110*W* | 0.500 | 0.0494 |
|  | 0.384 | 0.1055 |
|  | 0.116 | 0.4009 |
|  | 9.25 × 10$^{-5}$ | 1.1069 |
| *F*160*W* | 0.355 | 0.0494 |
|  | 0.543 | 0.1206 |
|  | 0.102 | 0.4117 |

*Note.* Circular MGE components. (1) Filter. (2) Normalization factor of Gaussian component (summing up to 1). (3) Standard deviation of Gaussian component.

**Table 7.** Parameters of the MGE luminosity conversion in the AB magnitude system.

| Target | Instrument | Filter | $M_{\odot, AB}$ (mag) | ZPT$_{AB}$ (mag) | Extinction (mag) |
| --- | --- | --- | --- | --- | --- |
| (1) | (2) | (3) | (4) | (5) | (6) |
| NGC 0997 | WFC3 | *F*160*W* | 4.60 | 25.94 | 0.071 |
| NGC 1684 | WFC3 | *F*110*W* | 4.52 | 26.82 | 0.051 |

*Note.* (1) Target name. (2) Instrument. (3) Filter. (4) Solar absolute magnitude. (5) Zero-point magnitude. (6) Galactic extinction.

for NGC 1684 and 0.071 mag for NGC 0997 from NED (Schlafly & Finkbeiner 2011), to correct our images for interstellar reddening. These parameters are summarized in Table 7. The parameters of the deconvolved best-fitting MGE Gaussians are listed in Table 8 in physical units and are illustrated in Fig. 5.

Given a (free) inclination, a 2D MGE surface brightness model can be analytically deprojected into a 3D light volume density distribution, which can itself be converted into a 3D stellar mass volume density distribution by simply multiplying each Gaussian component by a stellar *M/L*. For both galaxies we assume *M/L* to be free but radially constant, but alternative models will be considered when exploring potential systematic biases (Section 5.1).

The deprojection of a galaxy surface brightness is generally non-unique, but if one assumes axisymmetry, the MGE modelling yields

**Table 8.** Parameters of the deconvolved best-fitting MGE components of our models of NGC 1684 and NGC 0997.

| Target | $I'_\odot$ | $\sigma'$ | $q'$ |
| | ($L_\odot$ pc$^{-2}$) | (arcsec) | |
| (1) | (2) | (3) | (4) |
|---|---|---|---|
| NGC 1684 | 33 295 | 0.214 | 0.572 |
| (*F*110*W*) | 14 692 | 0.583 | 0.804 |
| | 8160 | 1.125 | 0.813 |
| | 3927 | 1.901 | 0.856 |
| | 2057 | 3.740 | 0.750 |
| | 737 | 9.255 | 0.666 |
| | 76 | 28.087 | 0.550 |
| | 147 | 28.087 | 0.900 |
| NGC 0997 | 13 772 | 0.056 | 0.941 |
| (*F*160*W*) | 25 967 | 0.300 | 0.914 |
| | 12 176 | 0.731 | 0.935 |
| | 6149 | 1.355 | 0.917 |
| | 937 | 2.310 | 0.922 |
| | 1492 | 4.951 | 0.892 |
| | 450 | 15.929 | 0.930 |

*Note.* Deconvolved MGE components. (1) Target name (filter). (2) Central surface brightness of Gaussian component. (3) Standard deviation of Gaussian component. (4) Axial ratio of Gaussian component.

a unique solution. While NGC 1684 is not perfectly axisymmetric, with a slight isophotal twist at large radii, it is reasonable to assume axisymmetry for our purpose of constraining the SMBH mass using gas kinematics (that primarily depends on the mass enclosed within small radii). NGC 0997, on the other hand, appears axisymmetric at all radii.

### 4.4 Fitting process

To identify the best-fitting model, we adopt a forward modelling approach whereby modelled data are compared against actual observations. For each galaxy, by adopting the gas distribution described in Section 4.2 and the circular velocity curve resulting from the axisymmetric mass distribution described in Section 4.3, the KinMS_fitter task of KINMS simulates the molecular gas disc as a collection of point-like sources. These sources are then used to create a model data cube by calculating their line-of-sight projections, taking into account the position (spatial centre and systemic velocity), inclination, and position angle (both assumed to be radially constant) of the galaxy (all free parameters of the model). Here, we assume the gas is in perfect circular rotation. This cube is then spatially convolved by the synthesized beam and binned into pixels identical to those of the real cube, to replicate instrumental effects.

Once a model data cube is generated, we compare it to the ALMA data cube using GASTIMATOR[5], a MCMC Gibbs-sampler with adaptive stepping that interfaces easily with KINMS. The MCMC samples the parameter space of all the free model parameters. For some of the parameters, we set reasonable priors (listed in Table 9) to ensure that the fitting process converges in a finite time. The other parameters are allowed to vary across their entire possible range. All priors are uniform linearly, except that on the SMBH mass which is uniform logarithmically because of its large dynamic range (although we ultimately revert to a linear sampling for NGC 0997, as explained in Section 5.2).

---

[5] https://github.com/TimothyADavis/GAStimator

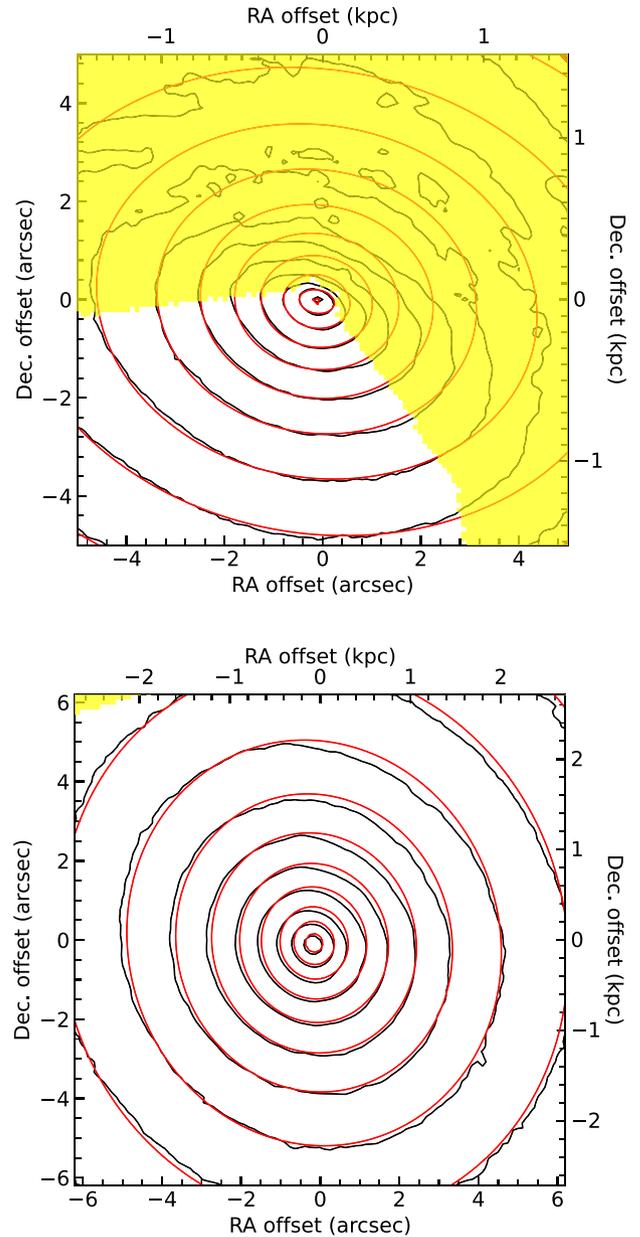

**Figure 5.** *Top:* *HST* WFC3 *F*110*W* image of NGC 1684 (black contours), overlaid with our best-fitting MGE model (red contours). The northern and western sides of the image are masked (yellow shading), to exclude the foreground dust disc and contamination from a nearby galaxy. **Bottom:** as the top panel, but for the *HST* WFC3 *F*160*W* image of NGC 0997. A small section of the north-eastern side of the image is masked, to exclude contamination from a nearby object. Both panels only show the small region of the galaxy relevant to our analysis.

Initially, the MCMC explores the parameter space in a manner such that the step size between each fit is adaptively scaled until the chain converges. Once the MCMC has converged, the maximum step size is fixed and the MCMC keeps sampling the parameter space, to generate samples from the final posterior probability distribution. The initial burn-in phase to identify the convergent chain usually comprises ≈ 10 per cent of the total number of steps. Assuming the maximally ignorant priors above and constant Gaussian errors throughout the cube, the posterior probability distribution of a model

Table 9. Best-fitting model parameters and associated uncertainties.

| Parameter | Prior | | | Best fit | $1\sigma$ uncertainty | $3\sigma$ uncertainty |
|---|---|---|---|---|---|---|
| **NGC 1684** | | | | | | |
| **Mass model** | | | | | | |
| log(SMBH mass/$M_\odot$) | 7 | $\to$ | 11 | 9.145 | $-0.048, +0.047$ | $-0.143, +0.118$ |
| Stellar $M/L_{F110W}$ ($M_\odot/L_{\odot, F110W}$) | 0.0 | $\to$ | 5.0 | 2.50 | $\pm 0.02$ | $\pm 0.05$ |
| **Molecular gas disc** | | | | | | |
| Inclination (°) | 0.0 | $\to$ | 89.9 | 66.7 | $\pm 0.3$ | $-0.8, +0.7$ |
| Position angle (°) | 0.0 | $\to$ | 359.9 | 258.1 | $\pm 0.2$ | $\pm 0.5$ |
| Velocity dispersion (km s$^{-1}$) | 0.0 | $\to$ | 50.0 | 13.3 | $-1.0, +1.1$ | $-2.7, +2.9$ |
| Scale length (arcsec) | 0.0 | $\to$ | 5.0 | 2.33 | $-0.09, +0.10$ | $-0.24, +0.31$ |
| $R_{\mathrm{hole}}$ (arcsec) | 0.0 | $\to$ | 3.0 | 0.39 | $-0.08, +0.07$ | $-0.37, +0.17$ |
| **Nuisance parameters** | | | | | | |
| Centre RA offset (arcsec) | $-3.2$ | $\to$ | 3.2 | 0.33 | $\pm 0.01$ | $\pm 0.03$ |
| Centre Dec. offset (arcsec) | $-3.2$ | $\to$ | 3.2 | 0.46 | $\pm 0.01$ | $\pm 0.02$ |
| Systemic velocity (km s$^{-1}$) | 4200 | $\to$ | 4500 | 4364.3 | $-1.0, +0.9$ | $-2.5, +2.4$ |
| Integrated intensity (Jy km s$^{-1}$) | 0.0 | $\to$ | 150.0 | 127.3 | $-5.8, +5.7$ | $-14.8, +15.8$ |
| **NGC 0997** | | | | | | |
| **Mass model** | | | | | | |
| SMBH mass ($10^8$ $M_\odot$) | 0 | $\to$ | 100 | 9.55 | $-3.81, +2.42$ | $-9.15, +8.87$ |
| Stellar $M/L_{F160W}$ ($M_\odot/L_{\odot, F160W}$) | 0.0 | $\to$ | 5.0 | 1.52 | $\pm 0.04$ | $\pm 0.11$ |
| **Molecular gas disc** | | | | | | |
| Inclination (°) | | (fixed) | | 34.3 | – | – |
| Position angle (°) | 0.0 | $\to$ | 359.9 | 215.2 | $\pm 0.5$ | $-1.3, +1.4$ |
| Velocity dispersion (km s$^{-1}$) | 0.0 | $\to$ | 50.0 | 10.2 | 1.1 | $-2.7, +3.4$ |
| **Nuisance parameters** | | | | | | |
| Centre RA offset (arcsec) | $-1.2$ | $\to$ | 1.2 | $-0.01$ | $\pm 0.02$ | $-0.05, +0.04$ |
| Centre Dec. offset (arcsec) | $-1.2$ | $\to$ | 1.2 | $-0.08$ | $\pm 0.02$ | $\pm 0.05$ |
| Systematic velocity (km s$^{-1}$) | 6100 | $\to$ | 6500 | 6357.2 | $-1.6, +1.7$ | $-4.1, +4.6$ |
| 2 arcsec $\times$ 2 arcsec integrated intensity (Jy km s$^{-1}$) | 0.0 | $\to$ | 50.0 | 9.4 | $-0.5, +0.6$ | $-1.5, +1.6$ |
| 12 arcsec $\times$ 12 arcsec integrated intensity (Jy km s$^{-1}$) | 0.0 | $\to$ | 50.0 | 44.2 | $-2.1, +2.2$ | $-5.5, +6.0$ |

*Note.* The RA and Dec. offsets are measured with respect to the image phase-centre. For NGC 1684 this is $04^h52^m31\overset{s}{.}166$, $-03°06'22\overset{''}{.}11$ (J2000.0); for NGC 0997 this is $02^h37^m14\overset{s}{.}485$, $+07°18'20\overset{''}{.}41$ (J2000.0).

is proportional to the log-likelihood function $\ln P \propto -0.5\chi^2$, where $\chi^2$ is defined in the standard manner as the sum (over all pixels of the cube) of the differences between model and data squared normalized by the uncertainties squared. We rescale the uncertainties of the cube by a factor of $(2N)^{0.25}$, where $N$ is the number of constraints (i.e. the number of pixels with detected emission, as defined by the mask in Section 3.1 and listed in Table 3), to ensure that the high $N$ does not lead to unrealistically small formal uncertainties. This method has been used in several other works (e.g. Mitzkus, Cappellari & Walcher 2017) and has been shown to yield uncertainties that are consistent with those found using a bootstrap approach (Smith et al. 2019).

Each of the free parameters of our model undergoes a process of marginalization as a part of the MCMC chain. The probability distributions of each parameter against the others are shown as 2D marginalization in Section 5, where each data point represents the log-likelihood of a model, the white data points being most likely and the blue data points least likely. The posterior probability distributions of the individual parameters are also shown as 1D marginalizations (i.e. histograms). All these histograms are roughly Gaussian in shape, indicating that the chains have converged.

Overall, our models have a total of nine basic free parameters: the SMBH mass ($M_{\mathrm{BH}}$), stellar mass-to-light ratio ($M/L$), integrated flux density, velocity dispersion of the molecular gas disc ($\sigma_{\mathrm{gas}}$), and the 'nuisance' disc parameters of position angle, inclination ($i$), systemic velocity ($V_{\mathrm{sys}}$), and centre position (in practice offsets in both right ascension and declination). When modelling the molecular gas distribution using a parametric surface density profile, as for NGC 1684, the exponential disc scale length ($R_0$) and truncation radius ($R_{\mathrm{hole}}$) are additional free parameters. The full list of model parameters, together with their search ranges, best-fitting values and $1\sigma$ and $3\sigma$ uncertainties, is provided in Table 9. The best-fitting models fit our data well, as will be discussed in Section 5.

## 5 RESULTS

### 5.1 NGC 1684

As shown in Fig. 5, there is significant dust extinction to the north of the NGC 1684 nucleus. To reduce their impact, the dust lanes were masked in our MGE model, simultaneously masking the nearby galaxy neighbour. The resulting MGE model is clearly a good fit, especially in the very centre of the image, the region that matters most to constrain the SMBH mass.

The final MCMC chain had 300 000 steps. It is clear that there is a large dark object present in the centre of NGC 1684, with a mass of $1.40^{+0.44}_{-0.39} \times 10^9$ $M_\odot$, where here and throughout this paper the uncertainties are stated at the $3\sigma$ (99.7 per cent) confidence level. The best-fitting *M/L* in the *F*110*W* filter is (2.50 ± 0.05) M$_\odot$/L$_{\odot, F110W}$. From Fig. 6, it is also clear that the well-known degeneracy between $M_{BH}$ and *M/L* is present, equivalent to the conservation of (total) dynamical mass.

The quality of our fit is easiest to judge by overlaying models over the innermost part of the kinematic major-axis PVD. From left to right in Fig. 7, we thus overlay the best-fitting models with no SMBH, the best-fitting SMBH, and an overly massive SMBH (0.3 dex more massive than the best fit), respectively, allowing all parameters other than the SMBH mass to vary. As expected for the best-fitting no SMBH model, a higher stellar *M/L* is derived to attempt to account for the high rotation velocities at small radii, but the model is nevertheless unable to fit those central velocities without greatly exceeding those relatively low velocities at larger radii. The compromise reached is thus unsatisfactory at both small and large radii (beyond the range shown in the figure). Again as expected, the best-fitting overly massive SMBH model yields a smaller *M/L*, but the fit is very poor at small radii, the model overshooting even the highest velocities. The best-fitting SMBH model not only fits the data best, but the fit is very good at all radii and velocities, including those radii beyond the range shown in the figure.

So far, we have only considered models with a radially constant *M/L*, but in principle this need not be the case. A spatially variable *M/L* can exist as a result of stellar population variations and it is worth exploring. Crucially, the issue is whether a good fit to the data can then be obtained without a SMBH.

First, we consider a model with a linear *M/L* by taking the circular velocity curve of an MGE model with an *M/L* of unity and scaling that curve by the square root of *M/L* (since $v^2 \propto M$). This adds one free parameter to our model. We ran the fit as before, allowing all parameters to vary. The best-fitting model has an *M/L* represented by $(M/L_{F110W})/\left(M_\odot/L_{\odot, F110W}\right) = 2.77 - 0.05 \, (R/\mathrm{arcsec})$. As shown in the left-hand panel of Fig. 8, the best-fitting no SMBH model with a linear *M/L* yields a marginally better fit at smaller radii than the best-fitting no SMBH model with a constant *M/L*, but a steep slope decreases the goodness of fit at larger radii and overall the fit is still very poor. Thus, an *M/L* linearly increasing inwards cannot replace the presence of a central SMBH.

Secondly, we consider a model where the innermost Gaussian of the MGE parametrization of the galaxy stellar mass has a different *M/L* than the other Gaussians. This effectively creates a model with a roughly Gaussian *M/L* at small radii that flattens out at larger radii. This adds one free parameter to our original model. We ran the fit as before, allowing all parameters to vary. The best-fitting model has an $M/L_{F110W}$ of 5.22 M$_\odot$/L$_{\odot, F110W}$ at the centre, 4.02 M$_\odot$/L$_{\odot, F110W}$ at a radius of 0.5 arcsec, and 2.95 M$_\odot$/L$_{\odot, F110W}$ at a radius of 1.0 arcsec, and the *M/L* flattens out at $\approx 2.6$ M$_\odot$/L$_{\odot, F110W}$ by a radius of $\approx 2.0$ arcsec. As shown in the right-hand panel of Fig. 8, this model does offer a good fit at both small and large radii, as a discrete increase of the *M/L* in the very centre effectively mimics a SMBH.

There is evidence suggesting that the centres of massive ETGs have a non-standard IMF (e.g. La Barbera et al. 2016; van Dokkum et al. 2017), with a mismatch parameter (the ratio between a galaxy's *M/L* and the predicted *M/L* assuming a Milky Way-like IMF) sufficient to give credence to our best-fitting model with a Gaussian *M/L* but no SMBH. However, most of the evidence comes from modelling of stellar populations or strong gravitational lenses, neither of which necessarily considers the effects of a central SMBH. Using a molecular gas method similar to ours, Davis & McDermid (2017) instead inferred a range of both radially increasing and decreasing dynamically determined *M/L*, implying both top- and bottom-heavy IMF variability between galaxies. Additionally, these gradients were significantly shallower than that of our best-fitting Gaussian *M/L* model. Thus, while a steep IMF gradient can technically explain the molecular gas kinematics of NGC 1684 without a central SMBH, given the established nature of central SMBHs, we do not seriously consider the model with a Gaussian *M/L* but no SMBH.

In fact, when inferring *M/L* dynamically, an appropriate treatment of the dynamics *should* take into account the gravitational effects of the SMBH, even when considering a variable *M/L*. When considering models with a SMBH and a linear *M/L*, our best-fitting model has a gradient of $\approx 0$. When considering models with a SMBH and a Gaussian *M/L*, a model with a central SMBH is strongly preferred over one without a SMBH. There is therefore no compelling reason to prefer a model with a variable *M/L* when accounting for the gravitational effects of the SMBH.

### 5.2 NGC 0997

As shown Fig. 5, there is no significant dust extinction close to the centre of NGC 0997, so significant masking near the sphere of influence of the SMBH is not required. The resulting MGE model is clearly a good fit, especially in the innermost region most important to the SMBH mass determination.

When modelling the whole extent of the galaxy, the SMBH mass remained poorly constrained, most likely due to the relatively small flux within $R_{SoI}$ compared to that within the whole galaxy. We thus conducted the modelling in two steps. For galaxy data sets for which constraining the inclination is problematic, it is a common practice to fix the inclination at a previously determined value (e.g. North et al. 2019). First, we therefore modelled the entire galaxy to obtain a robust inclination, using the entire molecular gas disc. Secondly, we restricted the modelling to a 2 arcsec × 2 arcsec central area, fixing the inclination to that obtained in the previous step. The restriction of the modelling to a central area is also common practice (e.g. Ruffa et al. 2023), especially when using SKYSAMPLER, which is computationally expensive. When possible, it is preferable to include the outer regions of the molecular gas disc in the modelling to obtain the most accurate *M/L* for the entire disc. However, in practice, a SMBH mass can be determined from modelling the central area of the galaxy only, where the SMBH dominates, and the inclusion of the outer regions is not strictly necessary.

The MCMC chain in the first stage modelling had 200 000 steps, yielding a best-fitting inclination $i = 34.3^{\circ +3.2}_{-2.6}$. Fig. 9 reveals a very strong degeneracy between *i* and *M/L*, but this is to be expected given the small inclination. Indeed, for a dynamical mass measurement based on rotational velocities, the enclosed mass $M_{tot}$ is related to the rotation velocity $v_{rot}$ and thus the observed velocity $v_{obs}$ as $M_{tot} \propto v_{rot}^2(R)R \propto v_{obs}^2(R)R/\sin^2 i \propto \sin^{-2} i$, and sin *i* varies rapidly at small inclinations (see Smith et al. 2019 for further discussion).

Initial modelling suggested the presence of a central SMBH in agreement with that predicted from the $M_{BH}$–$\sigma_e$ relation, but we were unable to decisively exclude very low ($<10^6$ M$_\odot$) SMBH masses. To treat this situation appropriately, we therefore chose to sample $M_{BH}$ directly (i.e. linearly), as opposed to sampling log ($M_{BH}$/M$_\odot$) (i.e. logarithmically) as done for NGC 1684. Indeed, when sampling the SMBH mass logarithmically, there is no natural lower boundary to the search, and $M_{BH} = 0$ can never be probed. When the posterior distribution extends all the way to the lower search boundary, the choice of prior thus affects the final distribution and in turn any upper limit adopted. By sampling the SMBH mass linearly, this arbitrariness can be removed by employing the least informative prior possible and probing all the way to 0. The final second-stage

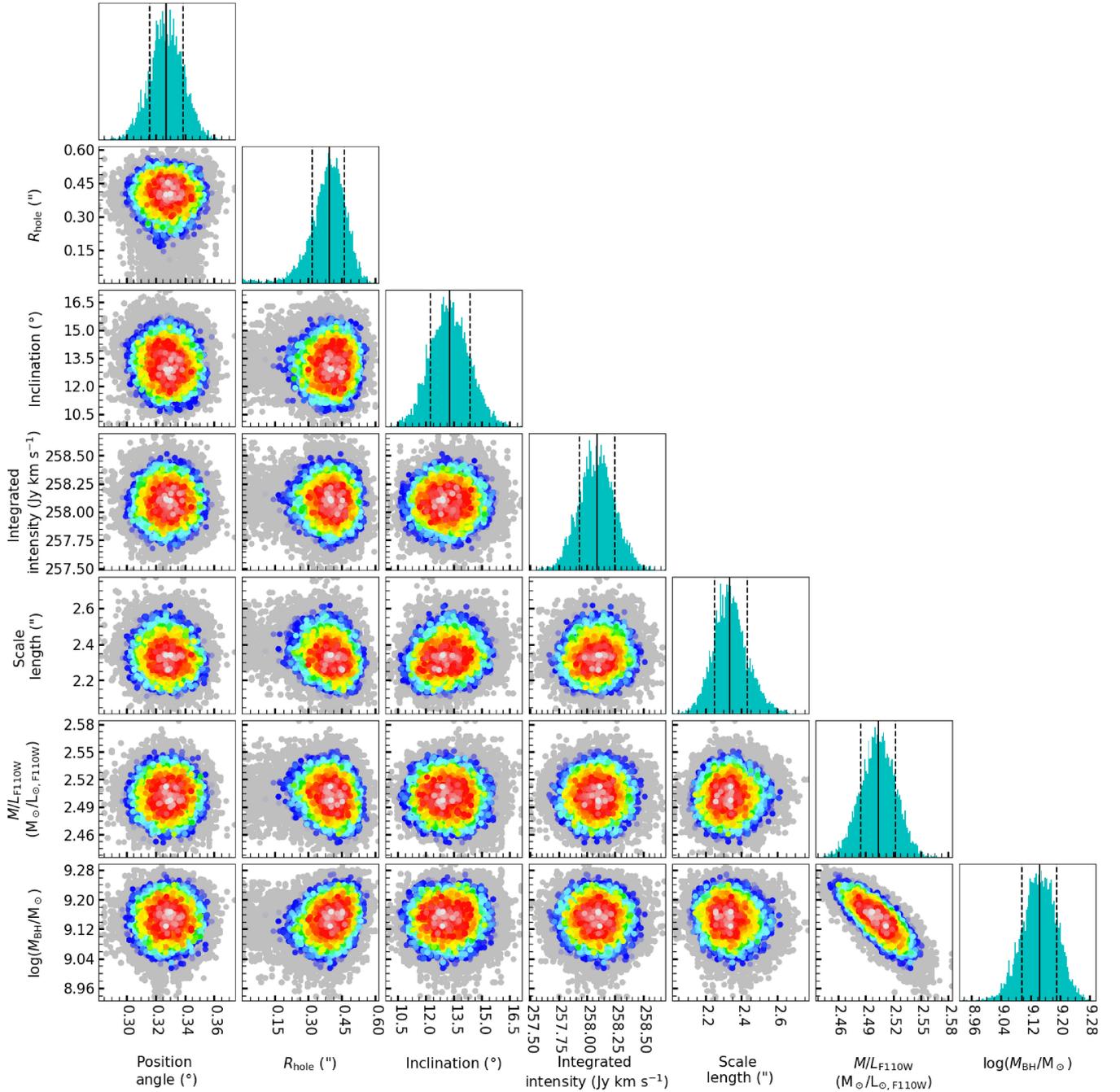

**Figure 6.** Corner plots of NGC 1684, showing the covariances between selected (primarily non-nuisance) model parameters. Each data point is a realization of our model, colour-coded to show the relative log-likelihood of that realization, with the white data points most likely and the blue data points least likely. The coloured points show models within $\Delta \chi^2 < \sqrt{2N}$ of the best-fitting model; the grey data points show all remaining models. The histograms show the 1D marginalized posterior distribution of each model parameter, where the black solid line marks the median and the two black dashed lines the 68 per cent confidence interval.

MCMC chain had 200 000 steps. Fig. 9 shows that there is a large dark object present in the centre of NGC 0997, with a mass ranging from $5.8 \times 10^8$ to $1.2 \times 10^9$ $M_\odot$ at the $1\sigma$ (68 per cent) confidence level, and from $4.0 \times 10^7$ to $1.8 \times 10^9$ $M_\odot$ at the $3\sigma$ (99.7 per cent) confidence level, with a best fit of $9.55 \times 10^8$ $M_\odot$. The best-fitting $M/L$ in the $F160W$ filter is $(1.52 \pm 0.11)$ $M_\odot/L_{\odot,F160W}$. As in NGC 1684, Fig. 9 shows the well-known degeneracy between $M_{BH}$ and $M/L$.

While the uncertainties on the SMBH mass of NGC 0997 are much larger than those of NGC 1684, and the SMBH mass posterior distribution superficially appears to extend all the way to 0, the lowest SMBH mass in the posterior distribution is actually $3 \times 10^6$ $M_\odot$, corresponding to a $4.3\sigma$ confidence interval. The presence of a SMBH is thus clearly necessary to reproduce the observed molecular gas distribution and kinematics. As before, the quality of the fit is easiest to judge by overlaying models over the innermost part of the kinematic major-axis PVD, so from left to right in Fig. 10 we overlay the best-fitting models with no SMBH, a free SMBH mass (equal to that of our best-fitting SMBH mass), and a fixed overly massive SMBH (≈0.3 dex more massive than the best fit), respectively, allowing all parameters other then the SMBH mass to vary. As expected, the best-fitting no SMBH model has a marginally larger $M/L$ but fails to accommodate the velocity enhancements within $R_{\mathrm{SoI}}$. Again as expected, the best-fitting model with an overly massive SMBH overshoots even the highest central velocities and yields a smaller $M/L$. Clearly, no satisfactory fit can be obtained with such a massive SMBH. The best-fitting SMBH model not only fits the data best and accounts very well for the central enhancements of velocities, but the fit is very good at all radii.

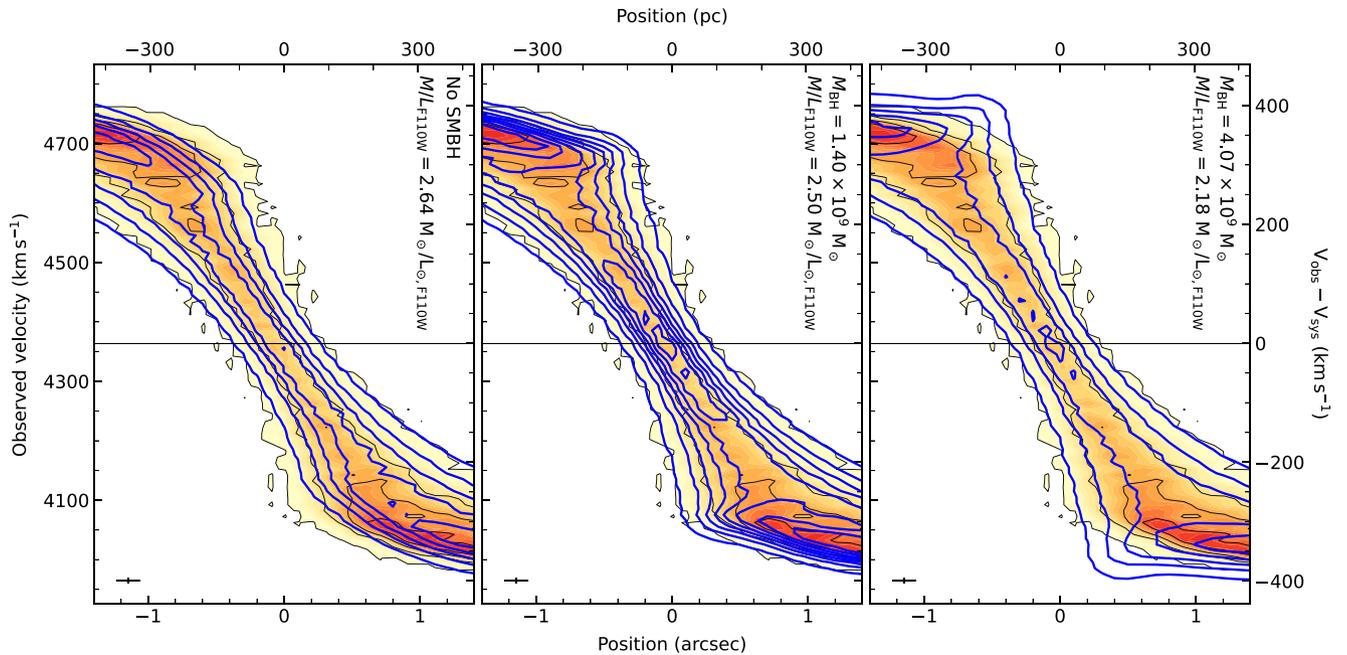

**Figure 7.** Innermost part (radii ≲1.4 arcsec) of the observed kinematic major-axis PVD of NGC 1684 (orange scale with black contours), overlaid with the best-fitting no SMBH (left), free SMBH mass (centre), and overly massive (by 0.3 dex) SMBH (right) model (blue contours), respectively. The SMBH mass and $M/L$ of each model are listed in the top-right corner of each panel. An error bar is shown in the bottom-left corner of each panel showing the size of the synthesized beam along the kinematic major-axis and the channel width. The need for a central dark mass to fully account for the gas kinematics at all radii is clear.

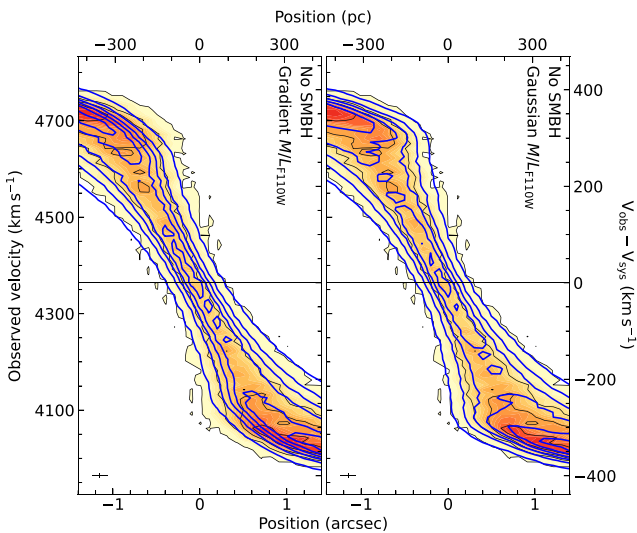

**Figure 8.** As Fig. 7, but for the best-fitting no SMBH with a linearly varying $M/L$ (left) and no SMBH with a Gaussian piecewise $M/L$ (right) model, respectively.

## 6 DISCUSSION

### 6.1 SMBH mass uncertainties

Aside from the fitting uncertainties reported in Table 9, SMBH mass measurements using molecular gas have several potential sources of systematic errors. Generally, the largest systematic error stems from the adopted distance. Indeed, as for all dynamical measurements, the SMBH mass scales linearly with distance, i.e. $M_{\mathrm{BH}} \propto D$, so any distance error impacts the inferred SMBH mass directly. The distances adopted here are those of the MASSIVE survey, and while no uncertainty is reported on the NGC 0997 distance measurement, it is generally accepted that the typical uncertainty on galaxy distances is ≈ 10 per cent, consistent with the formal errors on the NGC 1684 distance measurement. As is standard practice, we do not include the distance uncertainty in our final dynamical SMBH mass measurement, as our results can simply be scaled to any other adopted distance.

With the excellent angular resolutions of our ALMA data, the molecular gas discs of both of our targets are revealed to have central holes (see Section 3.1). The cumulative mass distribution of NGC 1684 shown in Fig. 11 shows that the synthesized beam size is much smaller than its hole, strongly suggesting the hole is real. This is supported by the posterior probability distributions shown in Fig. 6, even though the truncation radius is only loosely constrained. Using the standard definition of the SMBH sphere of influence (see Section 2) and our best-fitting SMBH mass, we derive $R_{\mathrm{SoI}} \approx 88$ pc (≈0.29 arcsec), while the best-fitting hole truncation radius $R_{\mathrm{hole}} = 0\farcs39^{+0.17}_{-0.37}$, slightly larger than $R_{\mathrm{SoI}}$ (see Fig. 11). This is most likely the reason we do not see a Keplerian rise of the rotation velocities near the galaxy centre. Fortunately, this does not prevent us from achieving a good constraint on the SMBH mass, as the SMBH influence extends beyond $R_{\rm SoI}$. In fact, the standard sphere of influence definition is only a rough estimate that often underestimates the true sphere of influence, defined as the radius where the enclosed stellar mass is equal to that of the SMBH. We will hereafter refer to this radius as the radius of equal mass contribution ($R_{eq}$). From Fig. 11, $R_{eq} \approx 105$ pc ($\approx 0.35$ arcsec), which is only slightly smaller than the hole. This most likely explains the robust inferred SMBH mass, as the molecular gas detected at the hole is still significantly affected by the SMBH.

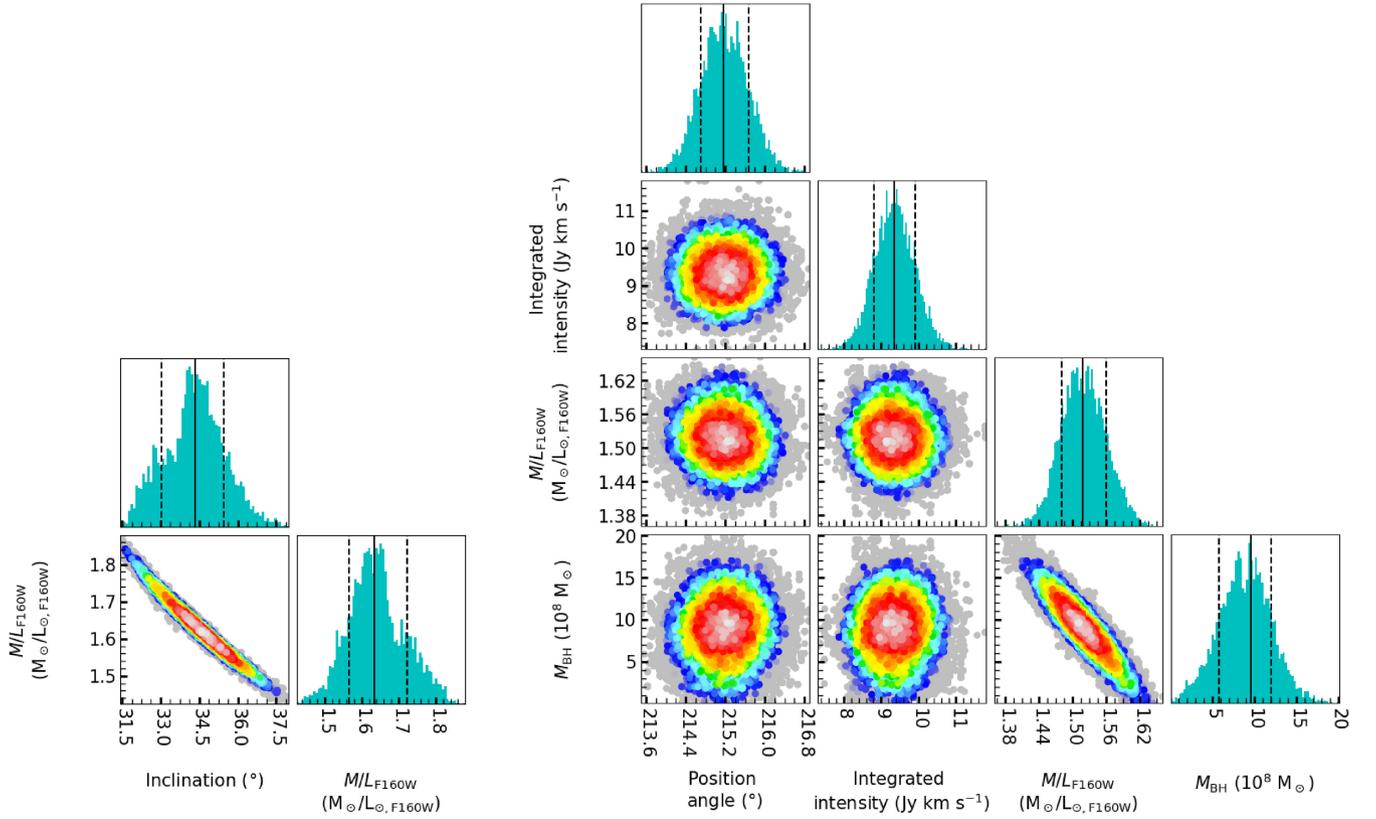

**Figure 9.** As Fig. 6, but for NGC 0997. **Left:** results from first modelling step, showing the covariance between *i* and *M/L*. **Right:** results from second modelling step, showing the covariances between all other non-nuisance parameters.

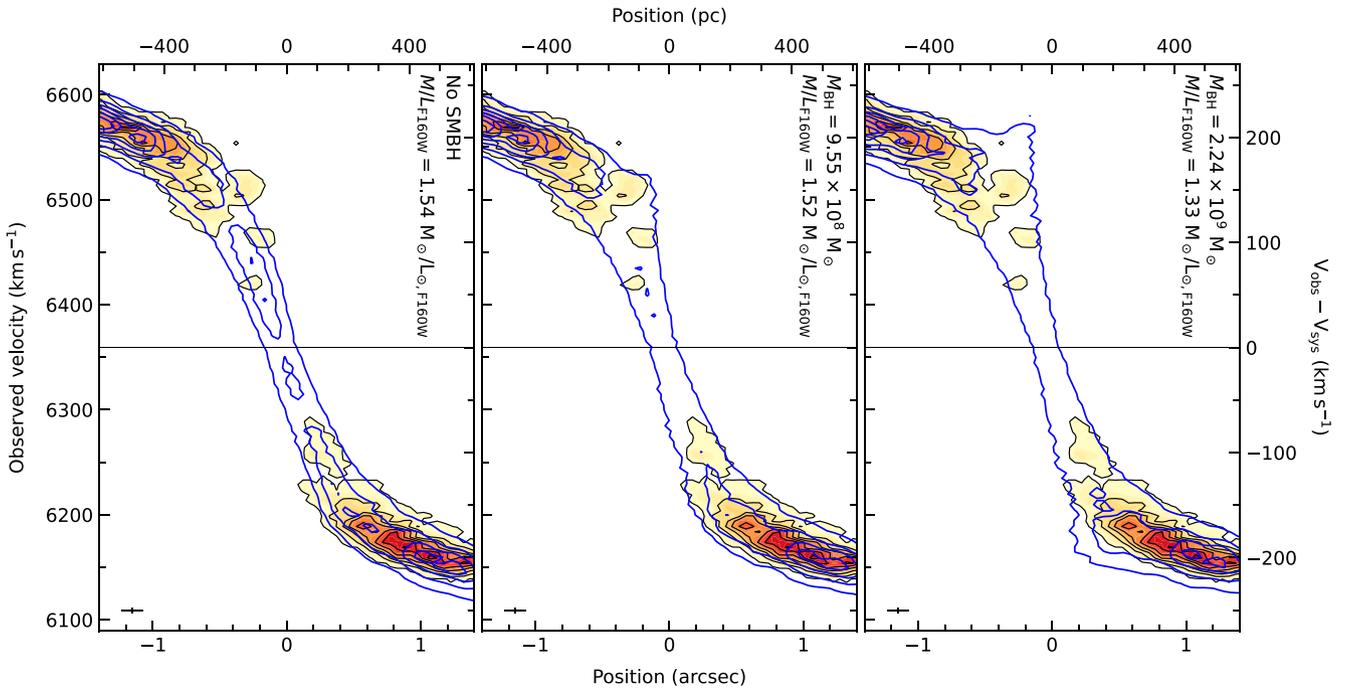

**Figure 10.** As Fig. 7, but for NGC 0997.

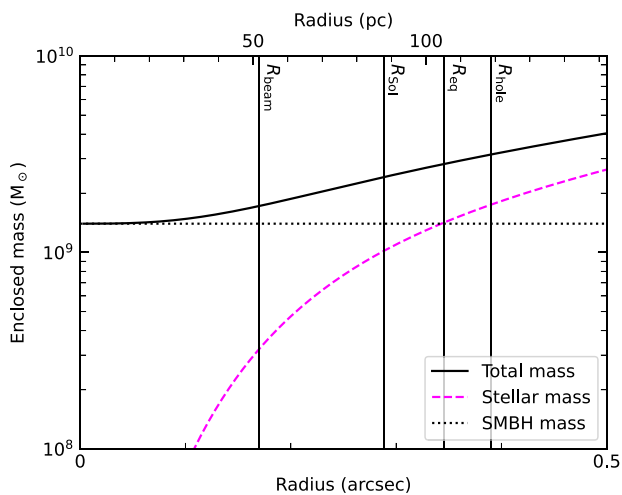

**Figure 11.** Cumulative mass function of NGC 1684, showing the relative contributions of the SMBH (black dotted line) and stars (magenta dashed line) to the total enclosed mass (solid black line). The four vertical black lines indicate the physical scales of the synthesized beam, the radius of the SMBH sphere of influence (assuming $\sigma_e = 262$ km s$^{-1}$ and our best-fitting SMBH mass), the radius of equal mass contribution, and our best-fitting hole truncation radius.

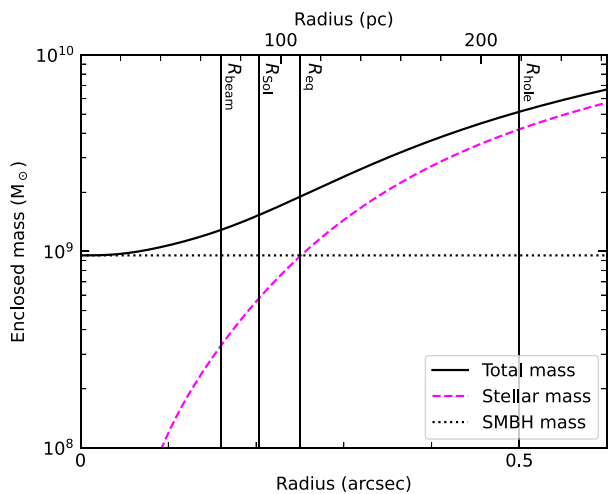

**Figure 12.** As Fig. 11, but for NGC 0997 (assuming $\sigma_e = 215$ km s$^{-1}$ and our best-fitting SMBH mass). We note that $R_{hole} \approx 0.50$ arcsec refers to the visually determined maximal extent of the asymmetric hole.

The situation of NGC 0997 is similar in some respects but different in others. The central molecular hole of NGC 0997 is asymmetric and non-circular, with a maximal extent of $R_{hole} \approx 0.50$ arcsec. The synthesized beam is again smaller than the hole, suggesting the hole is real. Using the standard definition of the SMBH sphere of influence and our best-fitting SMBH mass yields $R_{SoI} \approx 89$ pc ($\approx 0.20$ arcsec). From the cumulative mass distribution of NGC 0997 shown in Fig. 12, $R_{eq} \approx 110$ pc ($\approx 0.25$ arcsec). As discussed earlier, the

PVD has an asymmetric enhancement of velocities at $\approx 0.2$ arcsec, well within $R_{eq}$ and approximately at $R_{SoI}$. As such, this enhancement is likely the result of a SMBH and is the reason the SMBH mass can be constrained despite the otherwise depleted central molecular gas.

In fact, for a given data quality, the accuracy of the SMBH mass measurement should scale with the fraction of the enclosed mass due to the SMBH at the innermost radius probed. This radius is usually roughly the size of the synthesized beam ($R_{beam}$), but when a hole is present it is really the truncation radius of the hole. In NGC 1684, $\approx 80$ per cent of the enclosed mass at $R_{beam}$ is due to the SMBH, which would have made a SMBH mass measurement trivial had it not been for the central hole. That fraction drops to $\approx 45$ per cent at $R_{hole}$, but fortunately this is still sufficient to provide a robust SMBH mass determination. In NGC 0997, $\approx 74$ per cent of the mass enclosed at $R_{beam}$ is due to the SMBH, which again would have made a SMBH mass measurement trivial if not for the central hole. This fraction however drops to $\approx 20$ per cent at the maximal extent of the hole, which probably explains why the $3\sigma$ confidence intervals are larger than expected.

In our modelling, we have only considered the mass contributions of the stars and SMBH, and we have assumed the dark matter and molecular gas to have negligible masses. As discussed in Section 2, NGC 1684 has a total molecular gas mass of $1.58^{+0.75}_{-0.71} \times 10^9$ M$_\odot$, which although significant is spread over the entire extent of the molecular gas disc ($\approx 6$ arcsec or $\approx 1.8$ kpc in radius). Likewise, NGC 0997 has a total molecular gas mass of $1.82^{+1.30}_{-1.21} \times 10^9$ M$_\odot$, but this is again spread over its entire molecular gas disc ($\approx 5$ arcsec or $\approx 2.2$ kpc in radius). Considering in addition the large central holes in the molecular gas, the mass contribution from molecular gas in the central region of each galaxy is negligible. In fact, by definition, it is formally nil at the innermost radii probed ($R_{hole}$), and the cumulative mass functions of molecular gas would not be visible in Figs 11 and 12 even if they were plotted. Similarly, the spatial scales over which dark matter effects are significant are much greater than those considered here. In the very unlikely scenario that there is significant dark but diffuse matter in the central regions, the SMBH mass could be overestimated, but unless the distribution is significantly different from that of the stars, as mentioned earlier the dark matter is more likely to simply be assimilated with the stars and lead to a different $M/L$.

### 6.2 The $M_{BH}$–$\sigma$ relation: comparison to literature

As outlined in Section 1, SMBH mass measurements are often considered in the wider context of how correlations with other host galaxy properties inform our understanding of galaxy formation and co-evolution. To investigate how our results compare to other SMBH mass measurements, in Fig. 13 we compare them (blue data points) to the $M_{BH}$–$\sigma_e$ relation of van den Bosch (2016; grey data points) as well as literature molecular gas mass measurements (red data points; Davis et al. 2013b, 2017a, b, 2020; Onishi et al. 2015, 2017; Barth et al. 2016; Boizelle et al. 2019; Nagai et al. 2019; North et al. 2019; Ruffa et al. 2019, 2023; Smith et al. 2019, 2021; Nguyen et al. 2020, 2021, 2022; Cohn et al. 2021; Kabasares et al. 2022).

The NGC 1684 SMBH mass is slightly large for its velocity dispersion, as is NGC 0997, but both are well within the scatter of the relation. In fact, as the stellar velocity dispersion measurements of both galaxies lack formal uncertainties, both results could in fact be fully consistent with the $M_{BH}$–$\sigma_e$ relation of van den Bosch.

As noted by others (e.g. Lauer et al. 2007; McConnell et al. 2011; Thomas et al. 2016; North et al. 2019), some very massive galaxies host SMBHs that lie on the upper edge of the scatter of the $M_{BH}$–$\sigma_e$ relation, among the so-called 'overmassive' SMBH population. As discussed in Section 1, 12 of the $\approx$100 galaxies of the MASSIVE sample had an existing SMBH mass measurement prior to this work. Among them, four galaxies host SMBHs significantly above the mean $M_{BH}$–$\sigma_e$ relation (NGC 383, NGC 1600, NGC 3842, and NGC 4889). Taken together with the other MASSIVE galaxies that lie within the scatter of the relation, our new measurements suggest that the overmassive SMBHs previously found in some of these massive ETGs may in fact be fairly uncommon ($\lesssim$30 per cent). Further work by MASSIVE and other groups (e.g. Nguyen, Cappellari & Pereira-Santaella 2023), focusing on increasing the number of massive ETGs with direct SMBH mass measurements, will enable better constraint on the incidence of overmassive SMBHs and thus help understand the high-mass end of the $M_{BH}$–$\sigma_e$ relation.

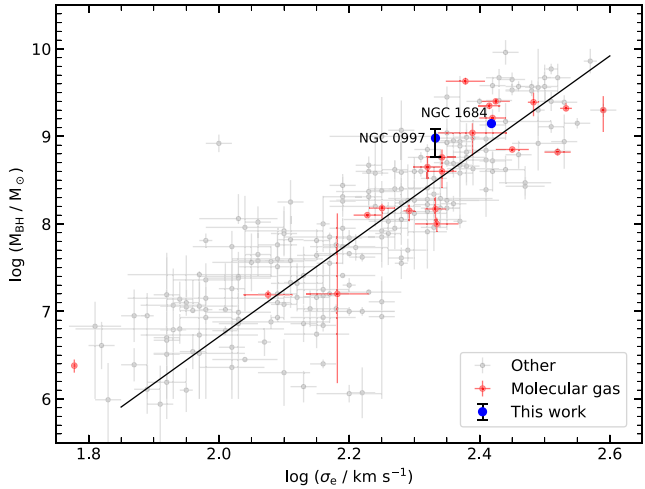

**Figure 13.** Black hole mass–stellar velocity dispersion relation obtained from stellar kinematic, ionized-gas kinematic, maser kinematic, and AGN reverberation mapping measurements (grey data points), as reported by van den Bosch (2016). The red data points show measurements obtained from molecular gas kinematics, while our measurements of the SMBH masses of NGC 1684 and NGC 0997 are shown in blue. Error bars denote $1\sigma$ uncertainties. The solid black line shows the best-fitting relation of van den Bosch (2016).

## 7 CONCLUSIONS

High angular resolution ALMA observations were obtained and used to make $^{12}$CO(2–1) cubes and 1.3 mm continuum images of the elliptical galaxies NGC 1684 and NGC 0997. We estimated the stellar mass distributions of both galaxies using MGE models of *HST* images and spatially constant $M/L$, and then forward modelled the molecular gas kinematics using KINMS and an MCMC framework. NGC 1684 has a regularly rotating molecular gas disc with a central hole of radius $0\rlap{.}{''}39^{+0.17}_{-0.37}$, preventing us from detecting a clear Keplerian rise of the rotational velocities at small radii. Nevertheless, we could infer a SMBH mass of $1.40^{+0.44}_{-0.39} \times 10^9$ M$_\odot$ ($3\sigma$ confidence level) and a stellar $F110W$ filter $M/L$ of $(2.50 \pm 0.05)$ M$_\odot$/L$_{\odot, F110W}$, very likely excluding the possibility of an alternative $M/L$ profile. NGC 0997 also has a regularly rotating molecular has disc, with an asymmetric central hole with a maximal radius of $\approx$0.5 arcsec, preventing us from detecting a Keplerian velocity rise. Nevertheless, we could infer a SMBH mass in the range $4.0 \times 10^7$–$1.8 \times 10^9$ M$_\odot$ and a stellar $F160W$ filter $M/L$ of $(1.52 \pm 0.11)$ M$_\odot$/L$_{\odot, F160W}$.

These SMBH measurements contribute to increasing the number of directly measured SMBH masses, thus allowing to constrain the $M_{BH}$–$\sigma_e$ relations over several orders of magnitude in mass, particularly here at the high-mass end. Our two measurements lie slightly above the mean $M_{BH}$–$\sigma_e$ relation but are consistent within the $\sim$0.4 dex intrinsic scatter of the relation (McConnell & Ma 2013; van den Bosch 2016), suggesting that some SMBHs of even very massive ETGs do follow the same relations.

## ACKNOWLEDGEMENTS

PD acknowledges support from a Science and Technology Facilities Council (STFC) DPhil studentship under grant ST/S505638/1. MB was supported by STFC consolidated grant 'Astrophysics at Oxford' ST/K00106X/1 and ST/W000903/1. This paper uses the following ALMA data: ADS/JAO.ALMA 2015.1.00187.S and 2016.1.0068.S. ALMA is a partnership of ESO (representing its member states), NSF (USA), and NINS (Japan), together with NRC (Canada), MOST and ASIAA (Taiwan), and KASI (Republic of Korea), in cooperation with the Republic of Chile. The Joint ALMA Observatory is operated by ESO, AUI/NRAO, and NAOJ. This research has used the NASA/IPAC Extragalactic Database (NED), which is operated by the Jet Propulsion Laboratory, California Institute of Technology, under contract with the National Aeronautics and Space Administration.

## DATA AVAILABILITY

The observations underlying this article are available in the ALMA archive, at https://almascience.eso.org/asax/, and in the Hubble Science Archive, at https://hst.esac.esa.int/ehst/.

## REFERENCES

Barth A. J., Boizelle B. D., Darling J., Baker A. J., Buote D. A., Ho L. C., Walsh J. L., 2016, ApJ, 822, L28
Boizelle B. D., Barth A. J., Walsh J. L., Buote D. A., Baker A. J., Darling J., Ho L. C., 2019, ApJ, 881, 10
Boizelle B. D. et al., 2021, ApJ, 908, 19
Bolatto A. D., Wolfire M., Leroy A. K., 2013, ARA&A, 51, 207
Cappellari M., 2002, MNRAS, 333, 400
Cappellari M., 2013, ApJ, 778, L2
Cohn J. H. et al., 2021, ApJ, 919, 77
Crook A. C., Huchra J. P., Martimbeau N., Masters K. L., Jarrett T., Macri L. M., 2007, ApJ, 655, 790
Dame T. M., 2011, preprint (arXiv:1101.1499)
Davis T. A., 2014, MNRAS, 443, 911
Davis T. A., McDermid R. M., 2017, MNRAS, 464, 453
Davis T. A. et al., 2013a, MNRAS, 429, 534
Davis T. A., Bureau M., Cappellari M., Sarzi M., Blitz L., 2013b, Nature, 494, 328
Davis T. A., Greene J., Ma C.-P., Pandya V., Blakeslee J. P., McConnell N., Thomas J., 2015, MNRAS, 455, 214
Davis T. A., Bureau M., Onishi K., Cappellari M., Iguchi S., Sarzi M., 2017a, MNRAS, 468, 4675
Davis T. A. et al., 2017b, MNRAS, 473, 3818
Davis T. A., Greene J. E., Ma C.-P., Blakeslee J. P., Dawson J. M., Pandya V., Veale M., Zabel N., 2019, MNRAS, 486, 1404
Davis T. A. et al., 2020, MNRAS, 496, 4061
Emsellem E., Monnet G., Bacon R., 1994, A&A, 285, 723
Ene I. et al., 2018, MNRAS, 479, 2810
Ene I., Ma C.-P., McConnell N. J., Walsh J. L., Kempski P., Greene J. E., Thomas J., Blakeslee J. P., 2019, ApJ, 878, 57
Ene I., Ma C.-P., Walsh J. L., Greene J. E., Thomas J., Blakeslee J. P., 2020, ApJ, 891, 65
Gebhardt K., Adams J., Richstone D., Lauer T. R., Faber S. M., Gültekin K., Murphy J., Tremaine S., 2011, ApJ, 729, 119
Goulding A. D. et al., 2016, ApJ, 826, 167

Goullaud C. F., Jensen J. B., Blakeslee J. P., Ma C.-P., Greene J. E., Thomas J., 2018, ApJ, 856, 11
Gu M., Greene J. E., Newman A. B., Kreisch C., Quenneville M. E., Ma C.-P., Blakeslee J. P., 2022, ApJ, 932, 103
Jensen J. B. et al., 2021, ApJS, 255, 21
Kabasares K. M. et al., 2022, ApJ, 934, 162
Kormendy J., Ho L. C., 2013, ARA&A, 51, 511
Krist J. E., Hook R. N., Stoehr F., 2011, in Kahan M. A., ed. Proc. SPIE Conf. Ser. Vol. 8127, Optical Modeling and Performance Predictions V. SPIE, Bellingham, 81270J
La Barbera F., Vazdekis A., Ferreras I., Pasquali A., Cappellari M., Martín-Navarro I., Schönebeck F., Falcón-Barroso J., 2016, MNRAS, 457, 1468
Lauer T. R., Tremaine S., Richstone D., Faber S. M., 2007, ApJ, 670, 249
Liepold C. M., Quenneville M. E., Ma C.-P., Walsh J. L., McConnell N. J., Greene J. E., Blakeslee J. P., 2020, ApJ, 891, 4
Liepold E. R., Ma C.-P., Walsh J. L., 2023, ApJ, 945, L35
Liu L., Bureau M., Blitz L., Davis T. A., Onishi K., Smith M., North E., Iguchi S., 2021, MNRAS, 505, 4048
Ma C.-P., Greene J. E., McConnell N., Janish R., Blakeslee J. P., Thomas J., Murphy J. D., 2014, ApJ, 795, 158
Magorrian J. et al., 1998, AJ, 115, 2285
McConnell N. J., Ma C.-P., 2013, ApJ, 764, 184
McConnell N. J., Ma C.-P., Gebhardt K., Wright S. A., Murphy J. D., Lauer T. R., Graham J. R., Richstone D. O., 2011, Nature, 480, 215
McConnell N. J., Ma C.-P., Murphy J. D., Gebhardt K., Lauer T. R., Graham J. R., Wright S. A., Richstone D. O., 2012, ApJ, 756, 179
McMullin J. P., Waters B., Schiebel D., Young W., Golap K., 2007, in Shaw R. A., Hill F., Bell D. J., eds, ASP Conf. Ser. Vol. 376, Astronomical Data Analysis Software and Systems XVI. Astron. Soc. Pac., San Francisco, p. 127
Mitzkus M., Cappellari M., Walcher C. J., 2017, MNRAS, 464, 4789
Nagai H. et al., 2019, ApJ, 883, 193
Nguyen D. D. et al., 2020, ApJ, 892, 68
Nguyen D. D. et al., 2021, MNRAS, 504, 4123
Nguyen D. D. et al., 2022, MNRAS, 509, 2920
Nguyen D. D., Cappellari M., Pereira-Santaella M., 2023, MNRAS, 526, 3548
North E. V. et al., 2019, MNRAS, 490, 319
Onishi K., Iguchi S., Sheth K., Kohno K., 2015, ApJ, 806, 39
Onishi K., Iguchi S., Davis T. A., Bureau M., Cappellari M., Sarzi M., Blitz L., 2017, MNRAS, 468, 4663
Pandya V. et al., 2017, ApJ, 837, 40
Pilawa J. D., Liepold C. M., Andrade S. C. D., Walsh J. L., Ma C.-P., Quenneville M. E., Greene J. E., Blakeslee J. P., 2022, ApJ, 928, 178
Quenneville M. E., Liepold C. M., Ma C.-P., 2022, ApJ, 926, 30
Quenneville M. E., Blakeslee J. P., Ma C.-P., Greene J. E., Gwyn S. D. J., Ciccone S., Nyiri B., 2023, MNRAS, 527, 249
Ruffa I. et al., 2019, MNRAS, 489, 3739
Ruffa I. et al., 2023, MNRAS, 522, 6170
Rusli S. P. et al., 2013, AJ, 146, 45
Sahu K. C., Anderson J., Baggett S., 2021, WFC3 Data Handbook, Version 5.0. STScI, Baltimore
Schlafly E. F., Finkbeiner D. P., 2011, ApJ, 737, 103
Shen J., Gebhardt K., 2010, ApJ, 711, 484
Skrutskie M. F. et al., 2006, AJ, 131, 1163
Smith M. D. et al., 2019, MNRAS, 485, 4359
Smith M. D. et al., 2021, MNRAS, 503, 5984
Thomas J., Ma C.-P., McConnell N. J., Greene J. E., Blakeslee J. P., Janish R., 2016, Nature, 532, 340
van den Bosch R. C. E., 2016, ApJ, 831, 134
van den Bosch R. C. E., Greene J. E., Braatz J. A., Constantin A., Kuo C.-Y., 2016, ApJ, 819, 11
van Dokkum P., Conroy C., Villaume A., Brodie J., Romanowsky A. J., 2017, AJ, 841, 68
Veale M. et al., 2017a, MNRAS, 464, 356
Veale M., Ma C.-P., Greene J. E., Thomas J., Blakeslee J. P., McConnell N., Walsh J. L., Ito J., 2017b, MNRAS, 471, 1428
Veale M., Ma C.-P., Greene J. E., Thomas J., Blakeslee J. P., Walsh J. L., Ito J., 2018, MNRAS, 473, 5446
Visser R., van Dishoeck E. F., Black J. H., 2009, A&A, 503, 323
Wada K., Fukushige R., Izumi T., Tomisaka K., 2018, ApJ, 852, 88
Willmer C. N. A., 2018, ApJS, 236, 47

## APPENDIX A: MOMENT MAPS

Here, we provide the moment maps for the observed data and best-fitting models for both galaxies.

## APPENDIX B: RESIDUALS

Here, we present the residuals generated by subtracting the first-moment (mean line-of-sight velocity field) of the best-fitting model cubes from that of the observed data cubes for both galaxies.

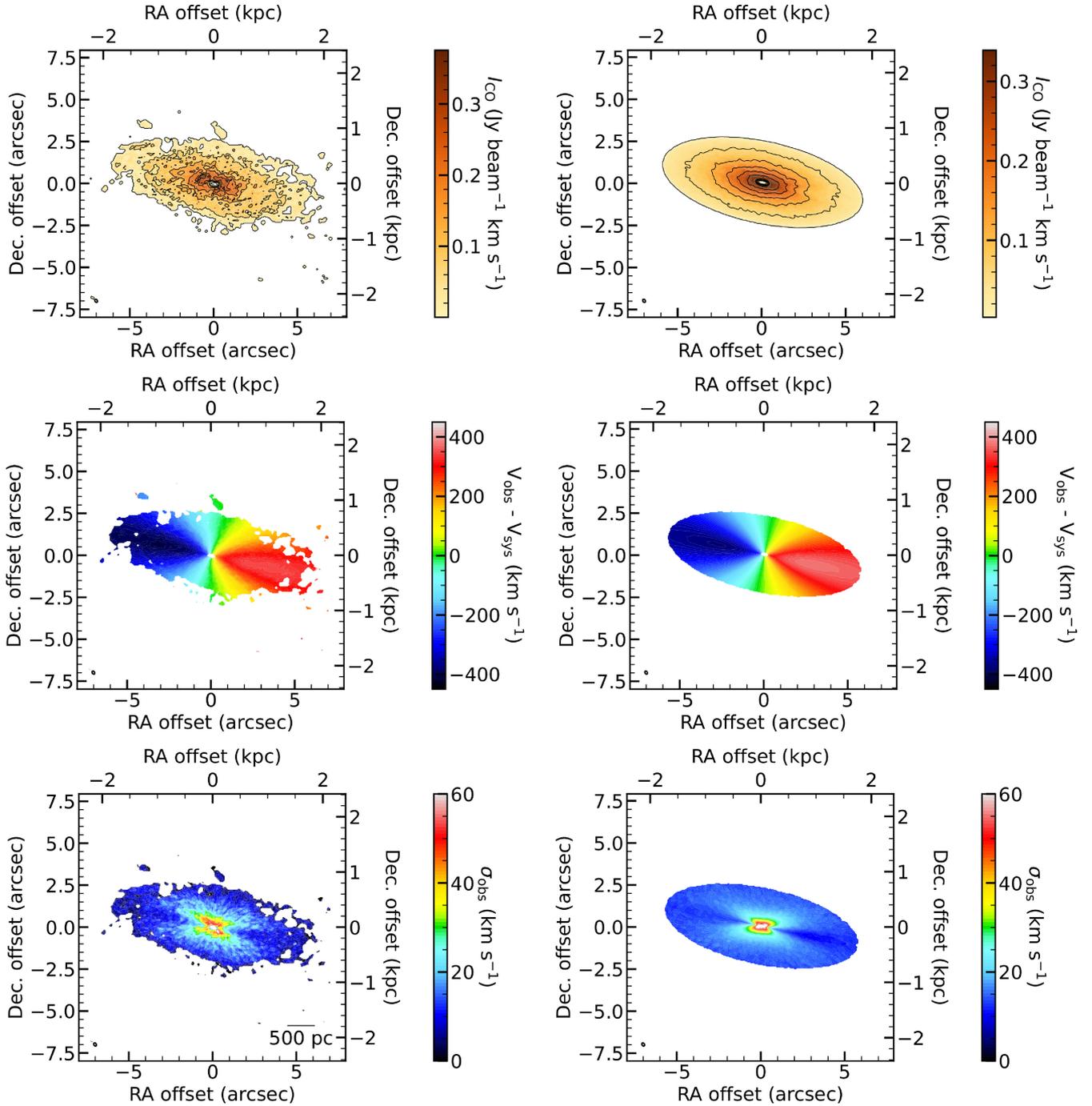

**Figure A1.** $^{12}$CO(2–1) zeroth-moment (integrated-intensity), first-moment (intensity-weighted mean line-of-sight velocity), and second-moment (intensity-weighted line-of-sight velocity dispersion) maps for NGC 1684. Moments extracted from observed data cubes are in the left-hand panels, while those extracted from the simulated best-fitting data cubes are in the right-hand panels.

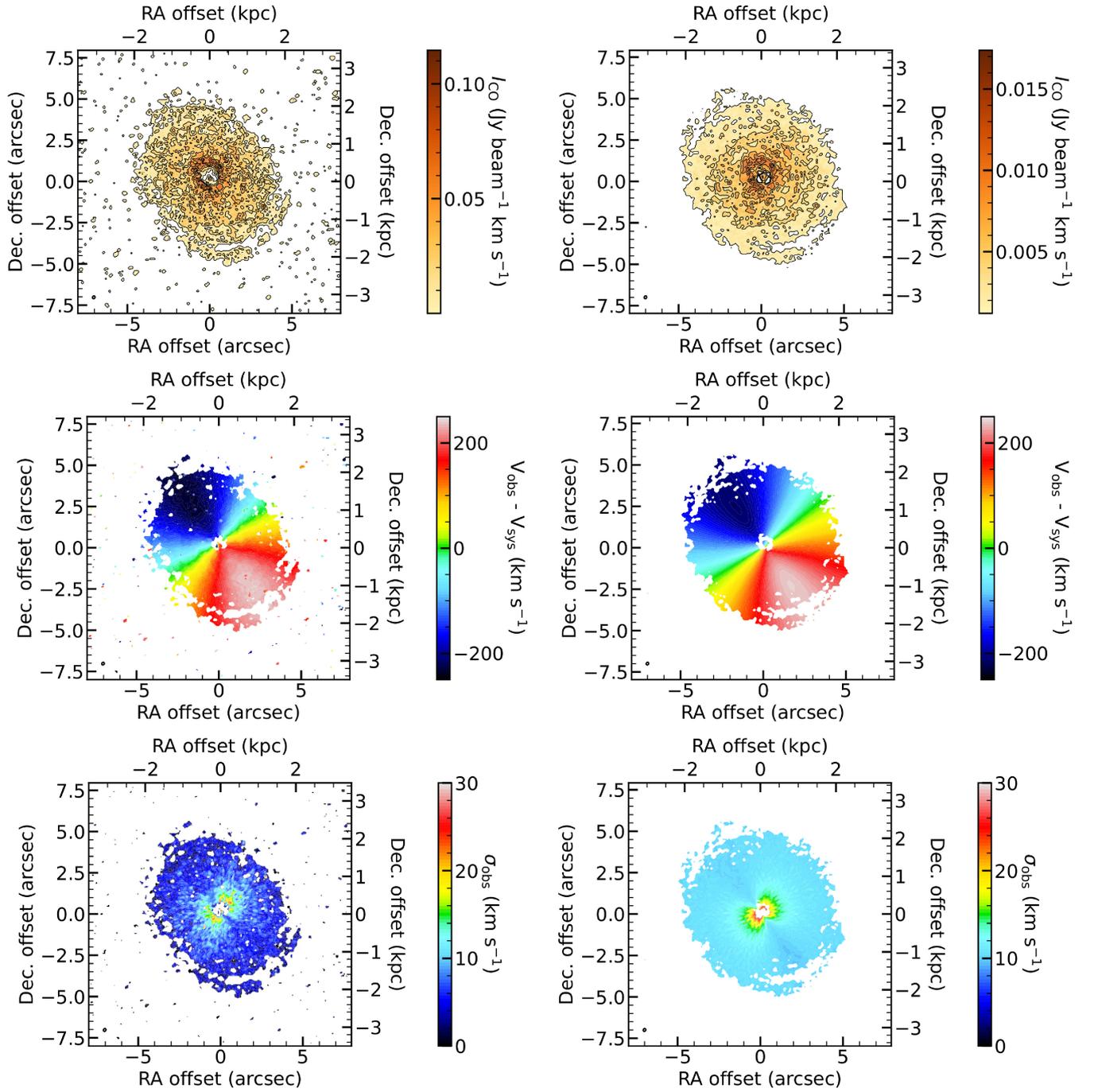

Figure A2. As Fig. A1 but for NGC 0997.

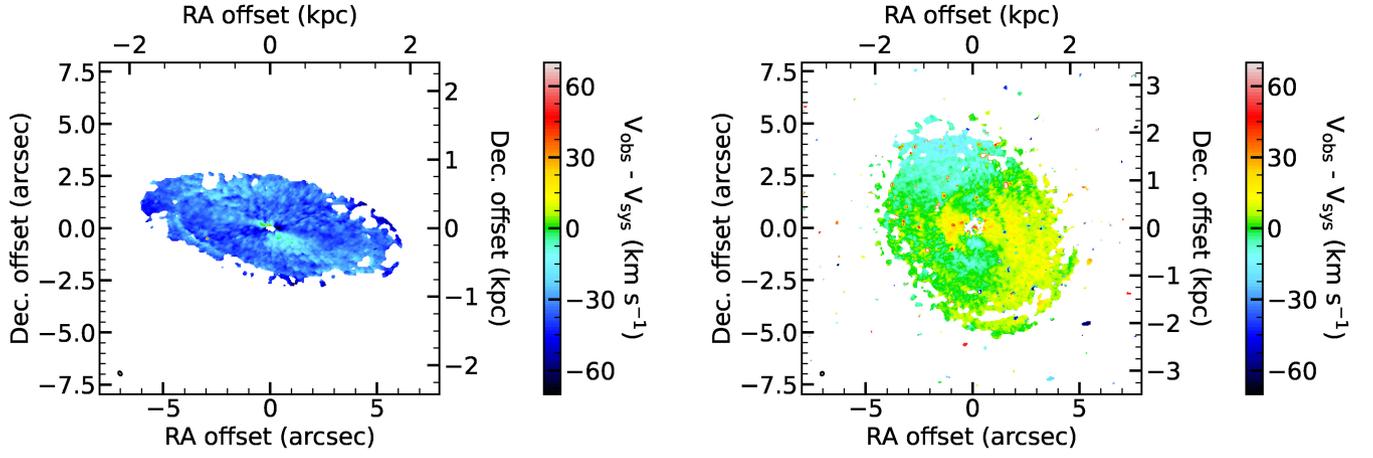

**Figure B1.** **Left:** Residual between the first-moments of the data cube and best-fitting model cube for NGC 1684. **Right:** As left-hand panel but for NGC 0997.

This paper has been typeset from a TEX/LATEX file prepared by the author.